%% file: HIG-17-002_temp.tex
\pdfoutput=1

\documentclass[11pt,twoside,a4paper,cmspaper,final,collab]{cms-tdr}
\def\svnVersion{444045P}\def\cmsCernNoTag{CERN-EP-2017-126}\def\cmsCernDate{\today}\def\cmsMessage{Published in Physics Letters B as \href{http://dx.doi.org/10.1016/j.physletb.2018.01.001}{\doi{10.1016/j.physletb.2018.01.001.}}}

\begin{document}\cmsNoteHeader{HIG-17-002}

\hyphenation{had-ron-i-za-tion}
\hyphenation{cal-or-i-me-ter}
\hyphenation{de-vices}
\RCS$Revision: 443973 $
\RCS$HeadURL: svn+ssh://svn.cern.ch/reps/tdr2/papers/HIG-17-002/trunk/HIG-17-002.tex $
\RCS$Id: HIG-17-002.tex 443973 2018-02-02 10:38:18Z lcadamur $
\newlength\cmsFigWidth
\ifthenelse{\boolean{cms@external}}{\setlength\cmsFigWidth{0.49\textwidth}}{\setlength\cmsFigWidth{0.6\textwidth}}
\ifthenelse{\boolean{cms@external}}{\providecommand{\cmsLeft}{upper\xspace}}{\providecommand{\cmsLeft}{left\xspace}}
\ifthenelse{\boolean{cms@external}}{\providecommand{\cmsRight}{lower\xspace}}{\providecommand{\cmsRight}{right\xspace}}
\newlength\cmsThreeWide\setlength{\cmsThreeWide}{0.32\textwidth}

\providecommand*{\bbtt}{\ensuremath{\PQb\PQb\PGt\PGt}\xspace}
\providecommand*{\tautau}{\ensuremath{\PGt\PGt}\xspace}
\providecommand*{\taue}{\ensuremath{\PGt_\Pe}\xspace}
\providecommand*{\taumu}{\ensuremath{\PGt_\PGm}\xspace}
\providecommand*{\muth}{\ensuremath{\PGt_\PGm\tauh}\xspace}
\providecommand*{\eleth}{\ensuremath{\PGt_\Pe\tauh}\xspace}
\providecommand*{\thth}{\ensuremath{\tauh\tauh}\xspace}
\providecommand*{\HH}{\ensuremath{\PH\PH}\xspace}
\providecommand*{\MTTwo}{\ensuremath{m_\text{T2}}\xspace}
\providecommand*{\MHHKinFit}{\ensuremath{m_{\HH}^\text{KinFit}}\xspace}
\providecommand*{\bbmth}{\ensuremath{\PQb\PQb\PGt_\PGm\tauh}\xspace}
\providecommand*{\bbeth}{\ensuremath{\PQb\PQb\PGt_\Pe\tauh}\xspace}
\providecommand*{\bbthth}{\ensuremath{\PQb\PQb\tauh\tauh}\xspace}
\providecommand*{\lambdahhh}{\ensuremath{\lambda_{\PH\PH\PH}}\xspace}
\providecommand*{\mX}{\ensuremath{m_\mathrm{S}\xspace}}
\providecommand*{\fb}{\unit{fb}\xspace}
\providecommand*{\myendash}{\mbox{--}}

\ifthenelse{\boolean{cms@external}}{\providecommand{\herebreak}{\relax}}{\providecommand{\herebreak}{\linebreak[4]}}
\cmsNoteHeader{HIG-17-002}
\title{Search for Higgs boson pair production in events with two bottom quarks and two tau leptons in proton-proton collisions at  $\sqrt{s} = 13\TeV$}

\date{\today}

\abstract{
A search for the production of Higgs boson pairs in proton-proton collisions at a centre-of-mass energy of 13\TeV is presented, using a data sample corresponding to an integrated luminosity of 35.9\fbinv collected with the CMS detector at the LHC. Events with one Higgs boson decaying into two bottom quarks and the other decaying into two \PGt leptons are explored to investigate both resonant and nonresonant production mechanisms. The data are found to be consistent, within uncertainties, with the standard model background predictions. For resonant production, upper limits at the 95\% confidence level are set on the production cross section for Higgs boson pairs as a function of the hypothesized resonance mass and are interpreted in the context of the minimal supersymmetric standard model. For nonresonant production, upper limits on the production cross section constrain the parameter space for anomalous Higgs boson couplings. The observed (expected) upper limit at 95\% confidence level corresponds to about 30\,(25) times the prediction of the standard model.
}

\hypersetup{%
pdfauthor={CMS Collaboration},%
pdftitle={Search for Higgs boson pair production in events with two bottom quarks and two tau leptons in proton-proton collisions at  sqrt(s) = 13 TeV},%
pdfsubject={CMS},%
pdfkeywords={CMS, physics, Higgs, Higgs boson pair production}}

\maketitle

\section{Introduction}
\label{sec:intro}

The discovery of the Higgs boson ($\PH$) by the ATLAS and CMS Collaborations~\cite{ATLASdiscovery,CMSdiscovery,Chatrchyan:2013lba}
was a major step towards improving the understanding of the mechanism of electroweak symmetry breaking (EWSB).
With the mass of the Higgs boson now precisely determined~\cite{Aad:2015zhl}, the structure
of the Higgs scalar field potential and the Higgs boson self-couplings are precisely predicted in the standard model (SM).
While the measured properties of the Higgs boson are thus far consistent with the expectations from the SM~\cite{Khachatryan:2016vau},
the measurement of the Higgs boson self-coupling provides an independent test of the SM and verification that the Higgs mechanism is truly responsible for the EWSB by giving access to the shape of the Higgs scalar field potential~\cite{Baglio:2012np}.

\begin{figure}[htb]
 \centering
 \includegraphics[width=\cmsFigWidth] {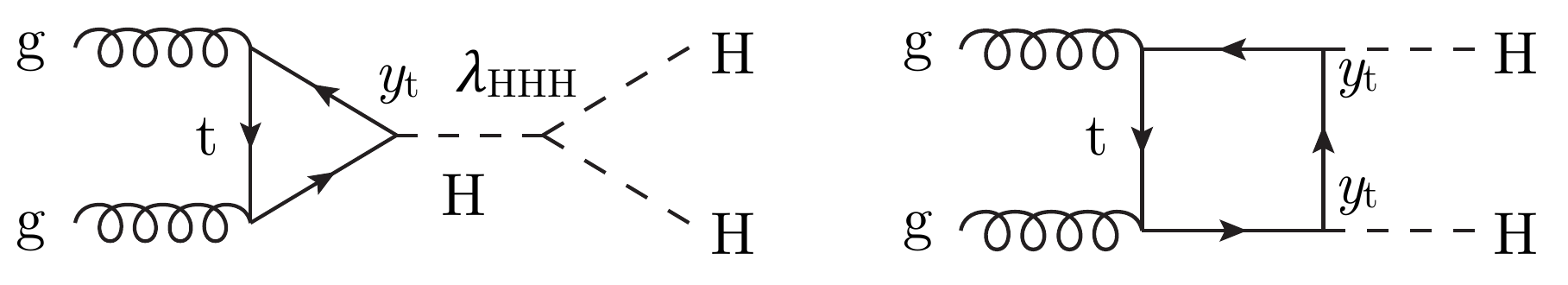}
 \caption{
 \label{fig:Feyndiag}
  Feynman diagrams contributing to Higgs pair production via gluon-gluon fusion at leading order at the LHC.}
 \end{figure}

The trilinear self-coupling of the Higgs boson (\lambdahhh) can be extracted from the measurement of the Higgs boson pair (\HH) production cross section.
In the SM, for proton-proton (pp) collisions at the CERN LHC, this process occurs mainly via gluon-gluon fusion and involves either couplings of the Higgs boson to virtual fermions in a quantum loop, or the \lambdahhh coupling itself, with the two processes interfering destructively as illustrated in Fig.~\ref{fig:Feyndiag}.

{\tolerance=1200
The SM prediction for the cross section is
$\sigma_{\PH\PH} = 33.49^{+4.3\%}_{-6.0\%}\, (\text{scale}) \pm 5.9\%\thy\fb$ \cite{deFlorian:2016spz,Borowka:2016ehy,Borowka:2016ypz,deFlorian:2015moa,Degrassi:2016vss}.
This value was computed at the next-to-next-to-leading order (NNLO) of the theoretical perturbative quantum chromodynamics (QCD) calculation, including
next-to-next-to-leading-logarithm (NNLL) corrections and finite top quark mass effects at next-to-leading order (NLO). The theoretical uncertainties in
$\sigma_{\PH\PH}$ include uncertainties in the QCD factorization and renormalization scales, the strong coupling parameter $\alpha_\text{S}$, parton distribution functions (PDF), and unknown effects from the finite top quark mass at NNLO.
\par}

Beyond the standard model (BSM) physics effects can appear either via anomalous couplings of the Higgs boson or via new particles that can be directly produced or contribute to the quantum loops responsible for \HH production.
The experimental signature would be an enhancement of the \HH production cross section
for a specific value of the invariant mass of the pair (resonant production) or over the whole invariant mass spectrum (nonresonant production).

Resonant double Higgs boson production is predicted by many extensions of the SM such as the singlet model~\cite{Binoth:1996au,Schabinger:2005ei,Patt:2006fw},
the two-Higgs-doublet model~\cite{Branco20121} and its realisation as the minimal supersymmetric standard model (MSSM)~\cite{Fayet:1974pd,Fayet:1977yc}, and models with warped extra dimensions (WED)~\cite{PhysRevD.76.036006, Fitzpatrick:2007qr}.
Although the physics motivation and the phenomenology of these theoretical models are very different, the signal is represented by a CP-even scalar particle (S)
decaying into a Higgs boson pair, with an intrinsic width that is often negligible with respect to the detector resolution.

In the nonresonant case, the BSM physics is modelled through an effective Lagrangian that extends the SM Lagrangian with
dimension-6 operators~\cite{Goertz:2014qta}.
Five Higgs boson couplings result from this parametrization: the Higgs boson coupling to the top quark, $y_\PQt$,
the trilinear coupling \lambdahhh, and three additional couplings, denoted as $\text{c}_2$, $\text{c}_{2\text{g}}$, and $\text{c}_\text{g}$ using the notation in Ref.~\cite{deFlorian:2016spz}, that represent, respectively, the interactions of a top quark pair with a Higgs boson pair, of a gluon pair with a Higgs boson pair, and of a gluon pair with a single Higgs boson.
For simplicity, we investigate only anomalous $y_\PQt$ and \lambdahhh couplings, while the other anomalous couplings are assumed to be zero, and parametrize the deviations from the SM values as $k_\lambda = \lambdahhh/\lambdahhh^\text{SM}$ and $k_\PQt = y_\PQt/y^\text{SM}_\PQt$.
Extension of these results to any combination of the couplings can be obtained by following the procedure detailed in Ref.~\cite{Carvalho:2015ttv}.
These two couplings are currently largely unconstrained by experimental results, and deviations from the SM can be accommodated by the combined measurements of Higgs boson properties~\cite{Khachatryan:2016vau} depending on the particular assumptions made about the BSM physics contributions.

Previous searches for the production of Higgs boson pairs were performed by both the
ATLAS~\cite{PhysRevD.92.092004, Aaboud:2016xco}
and CMS~\cite{PhysRevD.94.052012, Khachatryan2015560}
Collaborations using the LHC data collected at $\sqrt{s} = 8$ and 13\TeV.
The most sensitive upper limit at 95\% confidence level (CL) on HH production corresponds to 43 times the rate predicted by the SM and is obtained from the combination of the $\HH\to\bbbar\Pgg\Pgg$ and $\HH\to\bbbar\TT$ decay channels using data collected at $\sqrt{s} = 8\TeV$~\cite{HIG-15-013-arxiv}.

In this Letter we present a search for Higgs boson pair production in the final state where one Higgs boson decays to \bbbar and the other decays to \TT. For simplicity, we refer to this process as $\HH\to\bbtt$ in the following, omitting the quark and lepton charges.
This process has a combined branching fraction of 7.3\% for a Higgs boson mass of 125\GeV.  Its sizeable branching fraction, together with the relatively small background contribution from other SM processes, makes this final state one of the most sensitive to HH production.
Three final states of the \PGt lepton pair are considered: one of the two \PGt leptons is required to decay into hadrons and a neutrino (\tauh), while the other can decay either to the same final state, or into an electron (\taue) or a muon (\taumu) and neutrinos. Together, these three final states include about 88\% of the decays of the \tautau system and are the most sensitive ones for this search.
The data sample analyzed corresponds to an integrated luminosity of 35.9\fbinv collected in pp collisions at $\sqrt{s} = 13$\TeV.

The search described in this Letter improves on the previous $\HH\to\bbtt$ results~\cite{HIG-15-013-arxiv} by including final states with a leptonic \PGt decay, improving the event categorization, introducing multivariate methods for the background rejection, and optimizing the event and object selection for the LHC collisions at $\sqrt{s} = 13$\TeV.

\section{The CMS detector}
\label{sec:CMS}

The central feature of the CMS apparatus is a superconducting solenoid of 6\unit{m} internal diameter, providing a magnetic field of 3.8\unit{T}. Within the solenoid volume are a silicon pixel and strip tracker, a lead tungstate crystal electromagnetic calorimeter (ECAL), and a brass and scintillator hadron calorimeter (HCAL), each composed of a barrel and two endcap sections. Forward calorimeters extend the pseudorapidity coverage provided by the barrel and endcap detectors.
Muons are  detected in gas-ionization chambers embedded in the steel flux-return  yoke outside the solenoid.
Events of interest are selected using a two-tiered trigger system~\cite{Khachatryan:2016bia}. The first level, composed of custom hardware processors, uses information from the calorimeters and muon detectors to select events at a rate of around 100\unit{kHz} within a time interval of less than 4\mus. The second level, known as the high-level trigger, consists of a farm of processors running a version of the full event reconstruction software optimized for fast processing, and reduces the event rate to less than 1\unit{kHz} before data storage.
A more detailed description of the CMS detector, together with a definition of the coordinate system used and the relevant kinematic variables, including pseudorapidity $\eta$ and azimuthal angle $\varphi$, can be found in Ref.~\cite{Chatrchyan:2008zzk}.

\section{Modelling of physics processes}
\label{sec:BKG}

Simulated samples of resonant and nonresonant \HH production via gluon-gluon fusion are generated at leading order (LO) precision with \MGvATNLO~2.3.2~\cite{Alwall2014}. In the case of resonant production,
separate samples are generated for
mass values of the resonance ranging from 250 to 900\GeV.
In the case of nonresonant production, separate samples are generated for different values of the effective Lagrangian couplings, including the couplings predicted by the SM~\cite{Carvalho:2015ttv, Carvalho:2016rys}. In the latter case, an event weight determined as a function of the generated \HH pair kinematics is applied to these samples to model signals corresponding to additional points in the effective Lagrangian parametrization.

Backgrounds arising from $\PZ/\Pgg^{*}\to \ell\ell$ and $\PW\to \ell\nu_\ell$ in association with jets (with $\ell = \Pe, \PGm, \PGt$),
diboson ($\PW\PW$, $\PZ\PZ$, and $\PW\PZ$), and SM single Higgs boson production are simulated with \MGvATNLO 2.3.2 at LO with MLM merging~\cite{Alwall:2007fs}, while the single top and $\ttbar$ backgrounds are simulated at NLO precision with \POWHEG 2.0 \cite{Re:2010bp, Campbell2015}.
The NNPDF3.0 \cite{Ball:2014uwa} PDF set is used.
In order to increase the number of simulated events that satisfy the requirements detailed in Section~\ref{sec:reco}, the inclusive simulation of the $\PZ/\Pgg^{*}$ and $\PW$ processes is complemented by samples simulated in selected regions of multiplicity, flavour, and the transverse momentum scalar sum of the partons emitted at the matrix element level. Signal and background generators are interfaced with \PYTHIA 8.212~\cite{Sjostrand2015159} with the tune CUETP8M1~\cite{Khachatryan:2015pea} to
simulate the multiparton, parton shower, and hadronization effects. The simulated events include multiple overlapping hadron interactions as observed in the data.

The $\ttbar$, $\PZ/\Pgg^{*}\to \ell\ell$, $\PW\to \ell\nu_\ell$ and single top quark samples are normalized to their theoretical cross sections at NNLO precision~\cite{Czakon:2011xx, Li:2012wna, Kidonakis:2013zqa}, and the diboson samples are normalized to their cross section at NLO precision~\cite{Campbell:2011bn}.
The single Higgs boson production cross section is computed at the NNLO precision of the QCD corrections and at the NLO precision of electroweak corrections~\cite{deFlorian:2016spz,Brein:2012ne,Harlander:2013mla,Brein:2003wg,Altenkamp:2012sx}.

\section{Object reconstruction and event selection}
\label{sec:reconstruction}
\label{sec:reco}
In order to reconstruct an $\HH\to\bbtt$ candidate event, it is necessary to identify the \Pe, \PGm, and \tauh leptons, the jets originating from the two \PQb quarks, and the missing transverse momentum vector \ptvecmiss, defined as the projection onto the plane perpendicular to the beam axis of the negative vector sum of the momenta of all reconstructed particle-flow objects in an event. Its magnitude is referred to as \ptmiss.

The particle-flow (PF) event algorithm~\cite{Sirunyan:2017ulk} reconstructs and identifies each individual particle (PF candidate) with an optimized combination of information from the various elements of the CMS detector.
The momentum of the muons is obtained from the curvature of the corresponding track.
The energy of electrons is determined from a combination of the electron momentum at the primary interaction vertex, as determined by the tracker, the energy of the corresponding ECAL cluster, and the energy sum of all bremsstrahlung photons spatially compatible with originating from the electron track.
The energy of charged hadrons is determined from a combination of their momentum measured in the tracker and the matching ECAL and HCAL energy deposits, corrected for zero-suppression effects and for the response function of the calorimeters to hadronic showers. Finally, the energy of neutral hadrons is obtained from the corresponding corrected ECAL and HCAL energies.
Complex objects, such as \tauh, jets, and the \ptvecmiss vector, are reconstructed from PF candidates.
For each event, hadronic jets are clustered from PF candidates with the infrared and collinear safe anti-\kt algorithm~\cite{Cacciari:2008gp, Cacciari:2011ma}, operated with distance parameters of 0.4 and 0.8. These jets are denoted as ``AK4'' and ``AK8'' in the following. Leptons from \PQb hadron decays within a jet are considered as constituents by the algorithm.
The jet momentum is determined as the vectorial sum of all particle momenta in the jet, and is found in the simulation to be within 5 to 10\% of the true momentum over the whole \pt spectrum and detector acceptance. The invariant mass of AK8 jets is obtained by applying the soft drop jet grooming algorithm~\cite{Butterworth:2008iy, Larkoski2014}, that iteratively decomposes the jet into subjets to remove the soft wide-angle radiation and mitigates the contribution from initial state radiation, underlying event, and multiple hadron scattering. Jet energy corrections are derived from the simulation, and are confirmed with in situ measurements using the energy balance of dijet, multijet, $\gamma$+jet, and leptonic Z+jet events~\cite{Chatrchyan:2011ds,CMS:2016ljj}.
The PF components of the jets are used to reconstruct \tauh candidates using the hadrons plus strips algorithm~\cite{Khachatryan:2015dfa, CMS:2016gvn}, combining either one or three charged particle tracks with clusters of photons and electrons to identify the decay mode of the $\PGt$ lepton.

Events in the \bbmth (\bbeth) final state have been recorded using a set of triggers that require the presence of a single muon (electron) in the event.
The selected events are required to contain a reconstructed muon (electron)~\cite{Khachatryan:2015hwa, Chatrchyan:2012xi} of $\pt > 23 (27)\GeV$
and $\abs{\eta} < 2.1$ and a reconstructed \tauh candidate~\cite{Khachatryan:2015dfa} of $\pt > 20\GeV$ and $\abs{\eta} < 2.3$.
The muon (electron) candidate must satisfy the relative isolation requirement $I^\text{rel} < 0.15\,(0.1)$~\cite{Khachatryan:2015hwa, Chatrchyan:2012xi},
while the \tauh candidate must satisfy the ``medium'' working point of a multivariate isolation discriminant~\cite{Khachatryan:2015dfa}, that corresponds to a signal efficiency of about 60\% and a jet misidentification rate ranging between 0.1\% and 1\% depending on the jet $\pt$. The reconstructed tracks associated to the selected electron, muon, and \tauh candidates must be compatible with the primary pp interaction vertex of the event. Electrons and muons erroneously reconstructed as a \tauh candidate are rejected using discriminants based on the information from the calorimeters and muon detectors and on the properties of the PF candidates that form the \tauh candidate, as is detailed in~\cite{Khachatryan:2015dfa}.

A trigger requiring the presence of two $\tauh$ candidates is used to record events in the \bbthth final state.
The selected events must contain two reconstructed $\tauh$ candidates with ${\pt > 45\GeV}$ and $\abs{\eta} < 2.1$, that are required to pass the ``medium'' working point of the multivariate isolation discriminant and whose associated tracks must be compatible with the primary pp interaction vertex of the event.
The discriminants that suppress the contribution from prompt electrons and muons are applied to
both \tauh candidates as in the \bbmth and \bbeth final states.

For all three final states, the two selected \PGt leptons are required to have opposite electric charge.
Events containing additional isolated muons or electrons are rejected to reduce the $\PZ/\Pgg^{*}\to \ell\ell$ background contribution.

Events selected with the criteria described above (\muth, \eleth, \thth) are required to have
two additional AK4 jets with $\pt > 20\GeV$ and $\abs{\eta} < 2.4$.
In the case of \HH production via a resonance of mass 700\GeV or higher,
the two jets originating from the $\text{H}\to\PQb\PQb$ decay partially overlap due to the high Lorentz boost
of the Higgs boson, and are reconstructed at the same time as two separate AK4 jets and
as a single AK8 jet. To profit from this information, the event is classified as ``boosted'' if it contains at least one AK8 jet of invariant mass larger than 30\GeV and $\pt > 170\GeV$ that is composed of two subjets, each geometrically matched to one of the selected AK4 jets ($\Delta R (\text{AK4}, \text{subjet}) < 0.4$, where $\Delta R = \sqrt{\smash[b]{(\Delta\eta)^2 + (\Delta\varphi)^2}}$ denotes the spatial separation of the jet candidates). The event is classified as ``resolved'' if any of these requirements is not satisfied. This classification provides a clear separation of the signal topology  against the \ttbar background, where the two jets are typically more spatially separated and not reconstructed as a single AK8 jet. The AK8 jet mass requirement is applied to reject candidates resulting from a single quark or gluon hadronization or poorly reconstructed by the soft drop algorithm.

The combined secondary vertex~\cite{Chatrchyan:2012jua} algorithm is applied to the selected jets to identify those originating from a bottom quark
and reduce the contribution from the multijet background where jets are initiated by light quarks
or gluon radiation. Both the ``medium'' and the ``loose'' working points of the b tagging discriminant~\cite{CMS:2016kkf} are used in this search as described below.
The efficiency and rate of erroneous b jet identification are about 60\%\,(80\%) and 1\%\,(10\%) respectively for the ``medium'' (``loose'') working point.

Jets reconstructed in events classified as ``resolved'' are defined as b-tagged if they satisfy the ``medium'' working point of the b tagging algorithm. These events are classified into two groups according to the number of b-tagged jets: the group with at least two b-tagged jets
(2b) has the best sensitivity, and the group with exactly one b-tagged jet (1b1j) increases the signal
acceptance.
Both AK4 jets previously selected in the events classified as ``boosted'' are required to satisfy the ``loose'' working point of the b tagging discriminant.

\section{Signal regions and discriminating observables}
After the object selection and event classification, the kinematic information of the event is exploited to reduce
the contribution from background processes.
The invariant mass of the two $\PGt$ lepton candidates, $m_{\PGt\PGt}$, is reconstructed using a dynamic likelihood technique called SVfit~\cite{1742-6596-513-2-022035}
that combines the kinematics of the
two visible lepton candidates and the missing transverse momentum in the event.
The $\PQb\PQb$ invariant mass, $m_{\PQb\PQb}$, is estimated from the two selected jet candidates for ``resolved'' topologies and
from the invariant mass of the AK8 jet for ``boosted'' topologies.
In the ``resolved'' case, the events are required to satisfy the condition:
\begin{equation}
\frac{\left(m_{\PGt\PGt} - 116\GeV \right)^2}{\left( 35\GeV \right)^2} + \dfrac{\left(m_{\PQb\PQb} - 111\GeV \right)^2}{\left( 45\GeV \right)^2} < 1,
\end{equation}
where the values of 35 and 45\GeV are related to the mass resolution of the \tautau and $\PQb\PQb$ systems
and 116 and 111\GeV correspond to the position of the expected reconstructed 125\GeV Higgs boson peak in the $m_{\PGt\PGt}$ and
$m_{\PQb\PQb}$ distributions, respectively. The selection has been optimized for the SM \HH process to obtain a signal efficiency of approximately 80\% and a
background reduction of about 85\% in the most sensitive event categories. The $m_{\PQb\PQb}$ peak is shifted below the Higgs boson mass value because the momenta of neutrinos from \PQb hadron decays are not measured. This effect also prevents the SVfit algorithm from fully recovering the \tautau system mass value. In the ``boosted'' case the events are required to satisfy:
\begin{equation}
\begin{split}
80 < m_{\PGt\PGt} < 152\GeV, \\
90 < m_{\PQb\PQb} < 160\GeV.
\end{split}
\end{equation}

In addition to the previous requirements, a multivariate discriminant is applied to the events in the resolved categories of the \bbmth and \bbeth final states to identify and reject the \ttbar process, which is the most important source of background. The discriminant is built using the boosted decision tree (BDT)~\cite{Hocker:2007ht, friedman2001} algorithm that is trained on a combination of \muth and \eleth simulated signal and background events. The algorithm identifies the kinematic differences between the two processes and assigns to every selected event a number that defines its compatibility with a signal or background topology.
Two separate BDT trainings are performed to achieve an optimal performance for all the signal processes studied.

{\tolerance=1200
One training is performed using resonant signals with masses $\mX \leq 350\GeV$ as input.
Eight variables are used in the discriminant training because of their good separation between signal and background:
$\Delta\varphi(\text{H}_{\PQb\PQb},\text{H}_{\tautau})$,
$\Delta\varphi(\text{H}_{\tautau}, \ptvecmiss)$,
$\Delta\varphi(\text{H}_{\PQb\PQb}, \ptvecmiss)$,
$\Delta R(\PQb,\PQb) \, \pt(\text{H}_{\PQb\PQb})$,
$\Delta R(\ell,\tauh) \, \pt(\text{H}_{\tautau})$,
$m_\text{T} (\ell)$,
$m_\text{T} (\tauh)$,
and ${\Delta\varphi(\ell, \ptvecmiss)}$. Here $\ell$ refers to the selected muon or electron,
$\text{H}_{\PQb\PQb}$ and $\text{H}_{\tautau}$ denote the $\text{H}$ boson candidates reconstructed from the two jets and the two \PGt leptons, respectively, and
$m_\text{T} (\ell) = \sqrt{\smash[b]{ (\pt^\ell + \ptmiss)^2 - (\ptvec^\ell+\ptvecmiss)^2}}$
denotes the transverse mass of the selected lepton candidate, with a similar definition for $m_\text{T} (\tauh)$.
The $\Delta R$ separations of the two b quarks and of the two tau leptons are multiplied by the $\text{H}_{\PQb\PQb}$ and
$\text{H}_{\tautau}$ candidate \pt respectively to reduce their dependence on the $\mX$ hypothesis. All the selected variables contribute significantly to the discrimination achieved with the trained BDT. The same training is used both for the search for resonant \HH production up to $\mX = 350\GeV$ and for the search
for nonresonant \HH production.
No loss of performance is observed by using this training in comparison to a dedicated training on nonresonant signals.
Different selections on the BDT discriminant output are applied in the two searches to maximize the sensitivity:
these selections correspond to a rejection of the \ttbar background of approximately 90 and 70\% for the resonant
and nonresonant searches, respectively, for a signal efficiency ranging between 65 and 95\% depending on the
signal hypothesis considered.
\par}

A second training is performed on the resonant signals of mass $\mX > 350\GeV$. The variables used as inputs to this training
are the same as in the previous case, but replacing
$\Delta R(\PQb,\PQb) \, \pt(\text{H}_{\PQb\PQb})$ and
$\Delta R(\ell,\tauh) \, \pt(\text{H}_{\tautau})$
with
$\Delta R(\PQb,\PQb)$ and
$\Delta R(\ell,\tauh)$.
The selection on the BDT output is chosen to maximize the sensitivity and corresponds to a
rejection of the \ttbar background of approximately 90\% for a signal efficiency ranging between 70 and 95\% depending on the value of $\mX$.
In the case of the resonant search, the selections applied to the two BDT discriminants define low-mass (LM) and high-mass (HM) signal regions.

In the resonant search, the invariant mass of the two visible \PGt lepton decay products and the two selected $\PQb$ jets
is used to search
for a possible signal above the expected background event distribution.
In order to improve the resolution and to enhance the sensitivity of the analysis, the invariant mass
is reconstructed using a kinematic fit ($\MHHKinFit$) that is detailed in Ref.~\cite{Khachatryan:2015tha}.
The fit is based on the four-momenta of the $\PGt$ and $\PQb$ candidates and on the $\ptvecmiss$ vector in the event, and is
performed under the hypothesis of two 125\GeV Higgs bosons decaying into a bottom quark pair and a $\PGt$ lepton
pair. The use of the kinematic fit improves the resolution on $m_{\HH}$ by about a factor of two compared to the four-body invariant mass of the reconstructed leptons and jets.

The stransverse mass or \MTTwo variable is used in the search for a nonresonant signal.
This variable, originally introduced for supersymmetry searches involving invisible particles in the
final state~\cite{Lester:1999tx, 0954-3899-29-10-304} and later proposed for
\HH searches in \bbtt events~\cite{Barr2014308}, is used to reconstruct events where two equal mass particles are produced and each undergoes a two-body decay into a visible and an invisible particle.
The \MTTwo variable is defined as the largest mass of the parent particle that is compatible with the kinematic constraints of the event.
In the case of the \bbtt decay, where the dominant background is \ttbar production, the parent particle is interpreted as the top quark that decays into a bottom quark and a $\PW$ boson.
Following the description in Ref.~\cite{Barr2014308}, we denote with $\vec{b}$, $\vec{b}'$ the momenta of the two selected b jets and with $m_b$, $m_{b'}$ their invariant masses,
and we introduce the $\vec{c}$, $\vec{c}\,'$ symbols to denote the momenta of the other particles produced in the top quark decay corresponding to the measured leptons and the neutrinos. We also set $m_c = m^\text{vis} (\PGt_1)$ and
$m_{c'} = m^\text{vis} (\PGt_2)$, where $m^\text{vis}$ denotes the invariant mass of the measured leptons or \tauh. Under this notation, \MTTwo is defined as:
\ifthenelse{\boolean{cms@external}}{
\begin{multline}
\label{eq:mtt2formula}
\MTTwo \left(m_b, m_{b'}, \vec{b_\mathrm{T}}, \vec{b'_\mathrm{T}}, \vec{p_\mathrm{T}}^{\Sigma}, m_c, m_{c'} \right)\\
 =
\min_{\vec{c_\mathrm{T}} + \vec{c_\mathrm{T}}' = \vec{p_\mathrm{T}}^{\Sigma}}
\left\{\max \left( m_\mathrm{T}, m'_\mathrm{T} \right) \right\},
\end{multline}
}{
\begin{equation}
\label{eq:mtt2formula}
\MTTwo \left(m_b, m_{b'}, \vec{b_\mathrm{T}}, \vec{b'_\mathrm{T}}, \vec{p_\mathrm{T}}^{\Sigma}, m_c, m_{c'} \right) =
\min_{\vec{c_\mathrm{T}} + \vec{c_\mathrm{T}}' = \vec{p_\mathrm{T}}^{\Sigma}}
\left\{\max \left( m_\mathrm{T}, m'_\mathrm{T} \right) \right\},
\end{equation}
}
where the constraint in the minimization is over the measured lepton momenta and the missing transverse momentum, \ie
$\vec{p_\mathrm{T}}^{\Sigma} = \vec{p_\mathrm{T}}^\text{vis} (\PGt_1) + \vec{p_\mathrm{T}}^\text{vis} (\PGt_2) + \ptvecmiss$.
In Eq.~\eqref{eq:mtt2formula}, the transverse mass $m_\mathrm{T}$ is defined as
\begin{equation}
m_\mathrm{T} \left( \vec{b_\mathrm{T}}, \vec{c_\mathrm{T}}, m_b, m_c \right) =
\sqrt{m_b^2 + m_c^2 + 2\left( e_b e_c - \vec{b_\mathrm{T}}\cdot\vec{c_\mathrm{T}} \right)},
\end{equation}
and the ``transverse energy'' $e$ of a particle of transverse momentum $\pt$ and mass $m$ is defined as
\begin{equation}
e = \sqrt{m^2 + \pt^2}.
\end{equation}

We use the implementation in Ref.~\cite{Lester2015} to perform the minimization of Eq.~\eqref{eq:mtt2formula}.

The \MTTwo variable has a large discriminating power between the \HH signal and the \ttbar background, as it is bounded above by the top quark mass $m_\PQt$ for the irreducible background process $\ttbar \to \PQb\PQb\,\PW\PW \to \PQb\PQb \, \PGt\PGn_\PGt\PGt\PGn_\PGt$, while it can assume larger values for the \HH signal where the tau and the b jet do not originate from the same parent particle.
Detector resolution effects and other decay modes of the \ttbar system (\eg jets from the $\PW$ boson misidentified as \tauh) result in an extension of the tail of the \MTTwo distribution in \ttbar events beyond the $m_\PQt$ value.

\section{Background estimation}
\label{sec:BKGEstimation}
The main background sources that contaminate the signal region are \ttbar production, $\PZ/\Pgg^{*}\to \ell\ell$ production
and QCD multijet events.

The backgrounds from \ttbar, single top, single Higgs boson, W boson in association with jets, and diboson processes are estimated from simulation, as described in Section~\ref{sec:BKG}.

The $\PZ/\Pgg^{*}\to \ell\ell$ background contribution is estimated using the simulation, where the LO modelling of jet emission in the $\PZ/\Pgg^{*}$ process is known to be imperfect~\cite{Khachatryan:2016crw}.
Therefore, correction factors are calculated
using events containing two isolated, opposite-sign muons compatible with the $\PZ\to\PGm\PGm$ decay in association with two jets that satisfy similar invariant mass criteria as in the signal region.
This $\PZ$+2 jets sample is divided into three control regions according to the number of $\PQb$-tagged jets (0, 1, and 2) and three correction factors are derived for the $\PZ/\Pgg^{*}$ production in association with 0, 1, or $\geq$2
generator level jets initiated by b quarks, and applied in the signal regions.

The multijet background is determined from data in a jet-enriched region defined by requiring that the two selected \PGt lepton candidates have the same electric charge.
The yield is obtained from this same-sign (SS) region, where all the other selections are applied as in the signal region.
The events in this region are scaled by the ratio of opposite-sign (OS) to SS event yields obtained in a multijet-enriched region with inverted \PGt lepton isolation.
The contributions of other backgrounds, based on predictions from simulated samples, are subtracted in the OS and SS regions.
The shape of the multijet background is estimated using the events in an SS region with relaxed \PGt lepton isolation, after subtracting the other background contributions.

\section{Systematic uncertainties}
\label{sec:syst}

The effects of an imperfect knowledge of the detector response, discrepancies between simulation and data,
and limited knowledge of the background and signal processes are accounted for in the analysis as systematic uncertainties.
They are separately treated as ``normalization'' uncertainties or ``shape'' uncertainties; the first affect the number of expected events in the signal region, while the second affect
their distributions.

\subsection{Normalization uncertainties}

The following normalization uncertainties are considered:

\begin{itemize}
\item The integrated luminosity is known with an uncertainty of 2.5\%~\cite{CMS:2017sdi}.
This value is obtained from dedicated Van der Meer scans and the stability of detector response during the data taking.
The uncertainty is applied to the signal and to \ttbar, W+jets, single top quark, single Higgs boson, and diboson backgrounds, but it is not applied to the multijet and Z+jets backgrounds because they are estimated or corrected from data.
\item Electron, muon, and \tauh lepton trigger, reconstruction and identification
efficiencies are measured using $\Z\to\Pe\Pe$,
$\Z\to\PGm\PGm$, and
$\Z \to \PGt\PGt \to \tauh \PGn_\PGt \PGm \PGn_\PGm \PGn_\PGt$ events collected
at $\sqrt{s} = 13 \TeV$.
The corresponding uncertainties  are considered as uncorrelated among the final states and
are about 3\% for electrons, 2\% for muons, and 6\% for \PGt leptons.
\item The uncertainty in the knowledge of the \tauh energy scale
is about 3\% for each \tauh candidate~\cite{CMS:2016gvn}, and its impact on the overall normalization ranges from 3 to 10\%
depending on the process being considered. This effect is
fully correlated with a corresponding shape uncertainty in the distribution of \MTTwo and \MHHKinFit.
\item Uncertainties arising from the imperfect knowledge of the jet and b jet measured energy~\cite{Chatrchyan:2011ds} have an impact of about 2\% for the signal processes and 4\% for the backgrounds.
\item Uncertainties in the b tagging efficiency in the simulation are evaluated
as functions of jet \pt and $\eta$~\cite{CMS:2016kkf} and result in an average value of 2 to 6\% for the samples with genuine b jets in the final state.
\item  For the \ttbar process, the uncertainty in the normalization of the cross section is \herebreak$+$4.8\%/$-$5.5\%. For the W+jets, single top quark, diboson, and single Higgs backgrounds, uncertainties range from 1 to 10\%.
\item The uncertainties in the three correction factors derived in the control regions with 0, 1, and 2
b-tagged jets for the $\PZ/\Pgg^{*}\to \ell\ell$
background are propagated from the control regions to the signal region, taking into account the correlation between them, and amount to an uncertainty in the range 0.1--2.5\%
\item The uncertainty in the multijet background normalization is estimated by propagating the statistical uncertainties in the number of events used for its determination in the region with the
sign requirement inverted, as described in Section~\ref{sec:BKGEstimation}, and ranges between 5 and 30\% depending on the final state and category. Additional sources of systematic uncertainties were found to be negligible with respect to the statistical component given the number of events in the signal and control regions.
\item The uncertainties in the signal cross section arising from scale variations result in an uncertainty in its normalization of $+4.3\%$/$-6.0\%$ while effects from other theoretical uncertainties such as uncertainties on $\alpha_\text{s}$, PDFs and finite top quark mass effects at NNLO amount to a further 5.9\% uncertainty.
\end{itemize}
The systematic uncertainties are summarized in Table~\ref{tab:systs}.

\begin{table*}
\centering
\newcolumntype{x}{D{,}{\myendash}{1.2}}
\topcaption{\label{tab:systs}Systematic uncertainties affecting the normalization of the different processes.}
\begin{tabular}{lxl}
	\hline
		Systematic uncertainty & \multicolumn{1}{c}{Value} & Processes \\
		\hline
		Luminosity & \multicolumn{1}{c}{2.5\%} & all but multijet, $\PZ/\Pgg^{*}\to \ell\ell$ \\
		Lepton trigger and reconstruction & 2,6\% & all but multijet \\
		$\PGt$ energy scale & 3,10\% & all but multijet\\
		Jet energy scale & 2,4\% & all but multijet\\
		b tag efficiency & 2,6\% & all but multijet\\
		Background cross section & 1,10\% & all but multijet, $\PZ/\Pgg^{*}\to \ell\ell$ \\
		
		$\PZ/\Pgg^{*}\to \ell\ell$ SF uncertainty & 0.1,2.5\% & $\PZ/\Pgg^{*}\to \ell\ell$ \\
		Multijet normalization & 5,30\% & multijet \\[1.8ex]
		
		Scale unc. & \multicolumn{1}{c}{$+4.3$\%/$-6.0$\%} & signals\\
		Theory unc.  &\multicolumn{1}{c}{5.9\%} & signals\\
		
		\hline
\end{tabular}
\end{table*}

\subsection{Shape uncertainties}

The following shape uncertainties are considered:

\begin{itemize}
\item The shape uncertainty affecting the kinematic distribution in the simulation of the \ttbar background is
estimated by varying the top quark $\pt$ distribution according to the uncertainties in
differential $\pt$ measurements described in Ref.~\cite{Khachatryan:2016mnb}, and has an impact smaller than 1\% on the sensitivity of the measurement.
\item Uncertainties due to the limited number of simulated events or due to the statistical fluctuations of events in the multijet control region are taken into account.
These uncertainties are uncorrelated across bins in the individual template shapes and their inclusion has an impact on the sensitivity smaller than 7\%.
\item Uncertainties due to the \tauh and jet energy scales are taken into account and are fully correlated with the associated
normalization uncertainties. Uncertainties in the energy scales for other objects have negligible impacts on the simulated event
distributions and are not taken into account.
\end{itemize}

\section{Results}
\label{sec:results}
\begin{figure*}[p]
\centering
{\includegraphics[width=\cmsThreeWide] {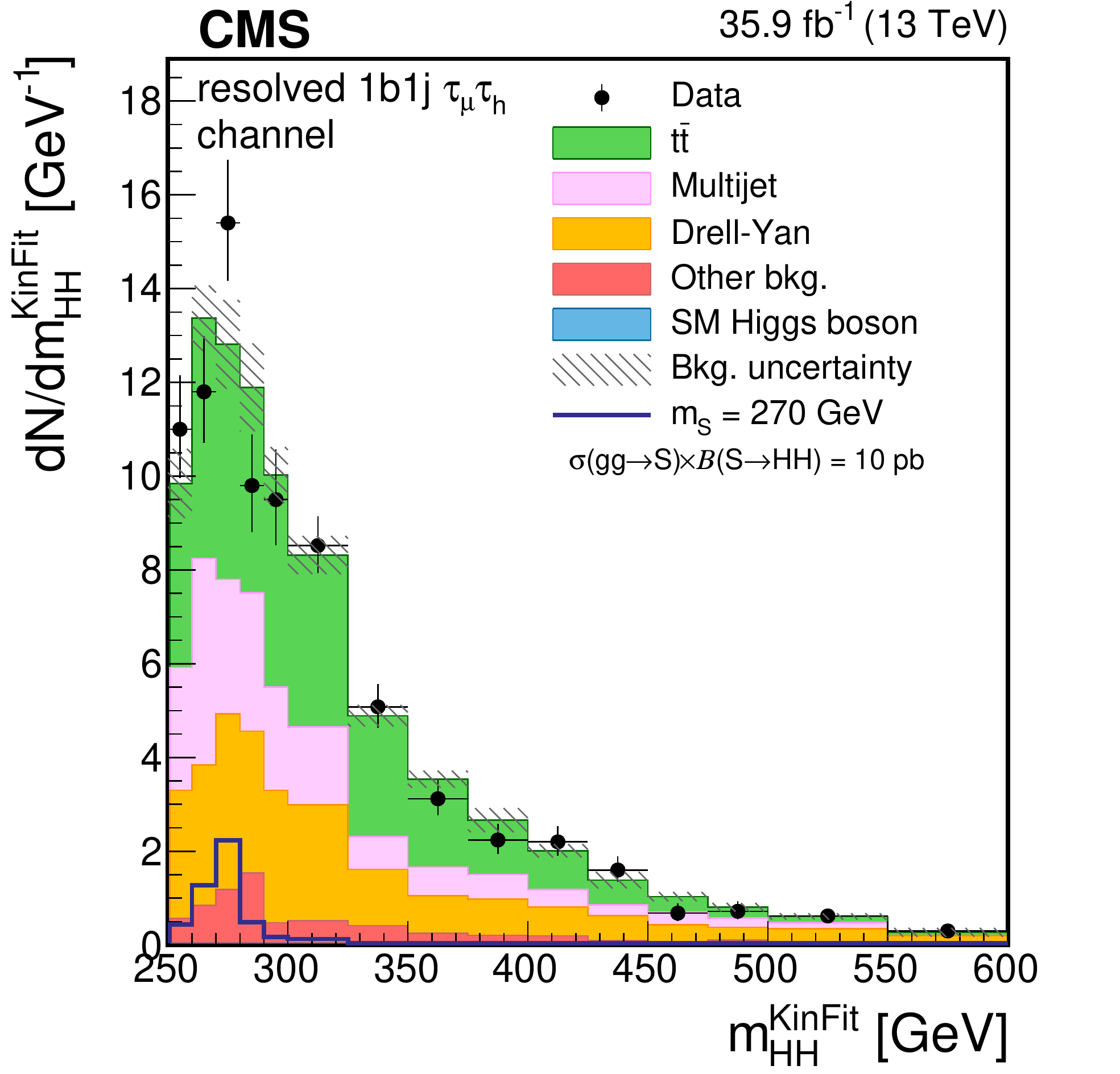}}\hspace{\fill}
{\includegraphics[width=\cmsThreeWide] {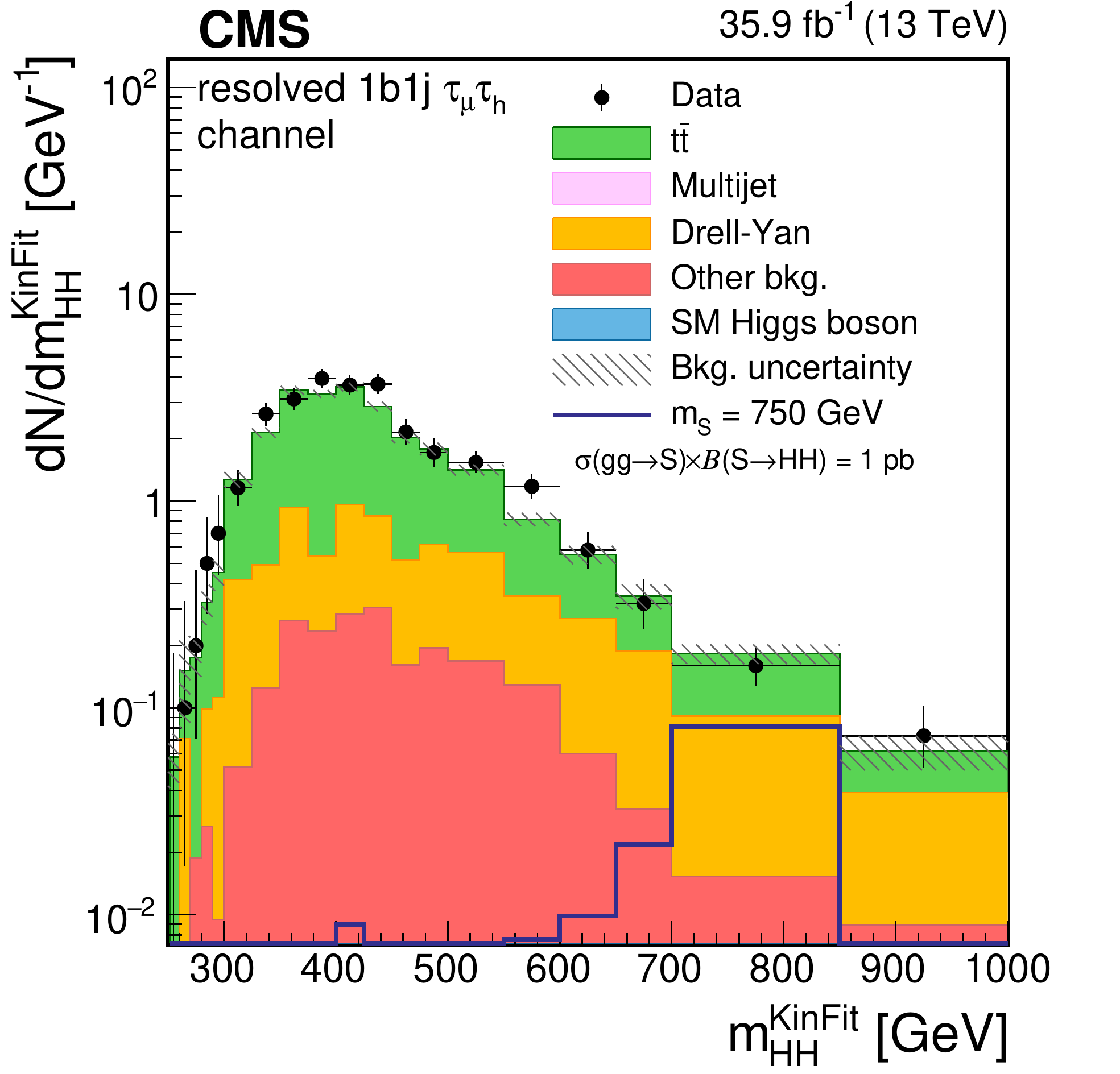}}\hspace{\fill}
{\includegraphics[width=\cmsThreeWide] {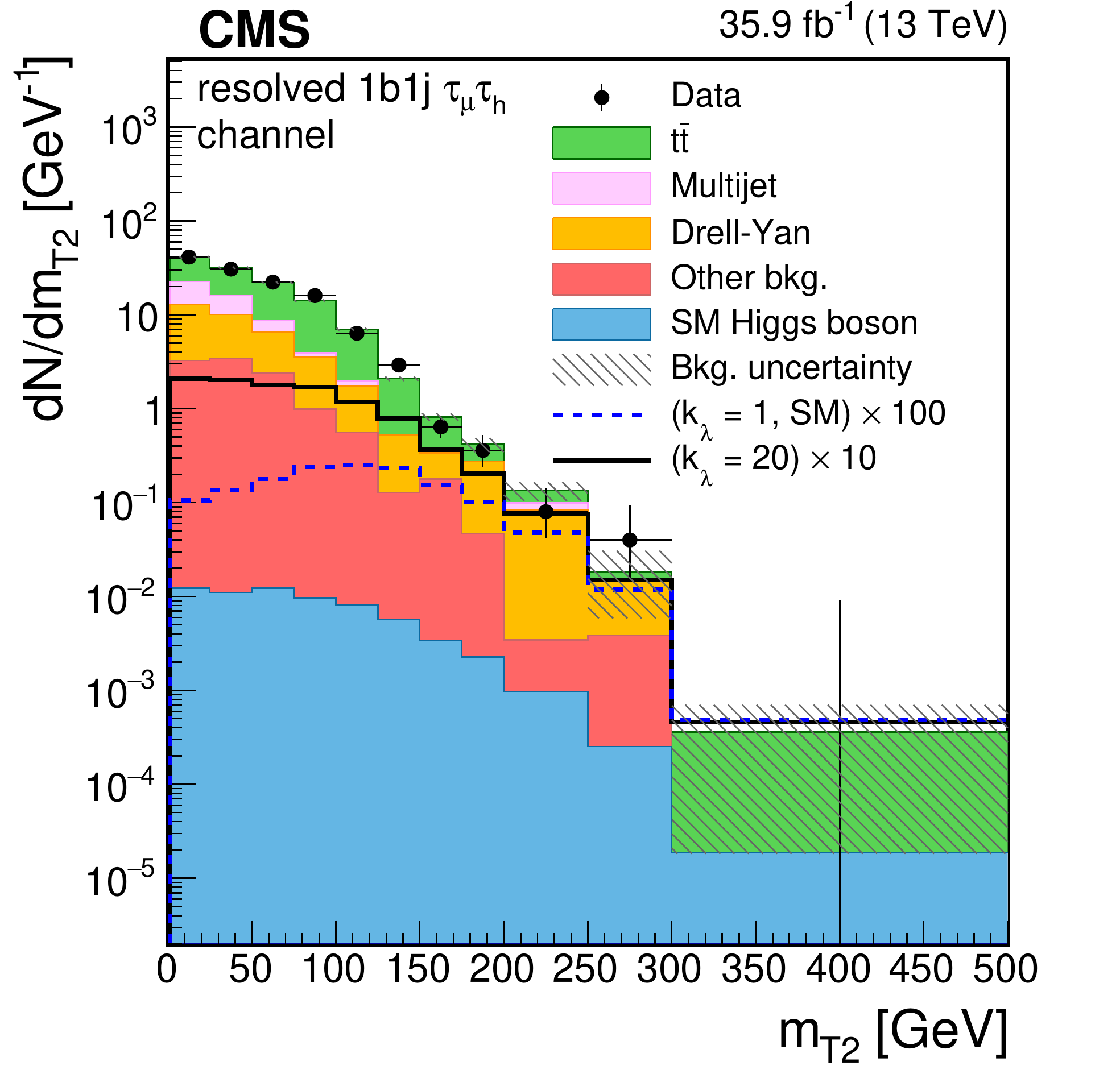}}\\
{\includegraphics[width=\cmsThreeWide] {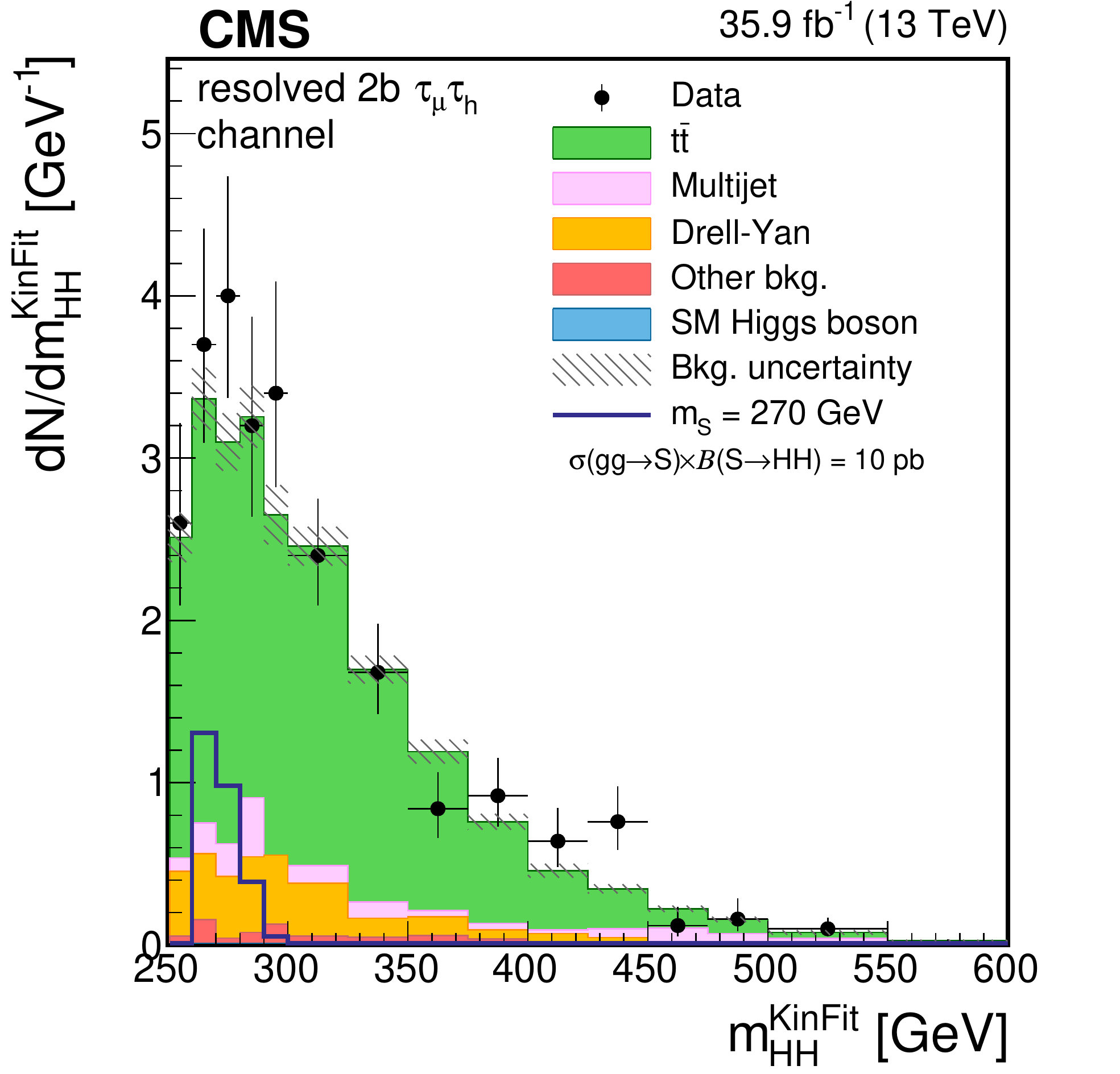}}\hspace{\fill}
{\includegraphics[width=\cmsThreeWide] {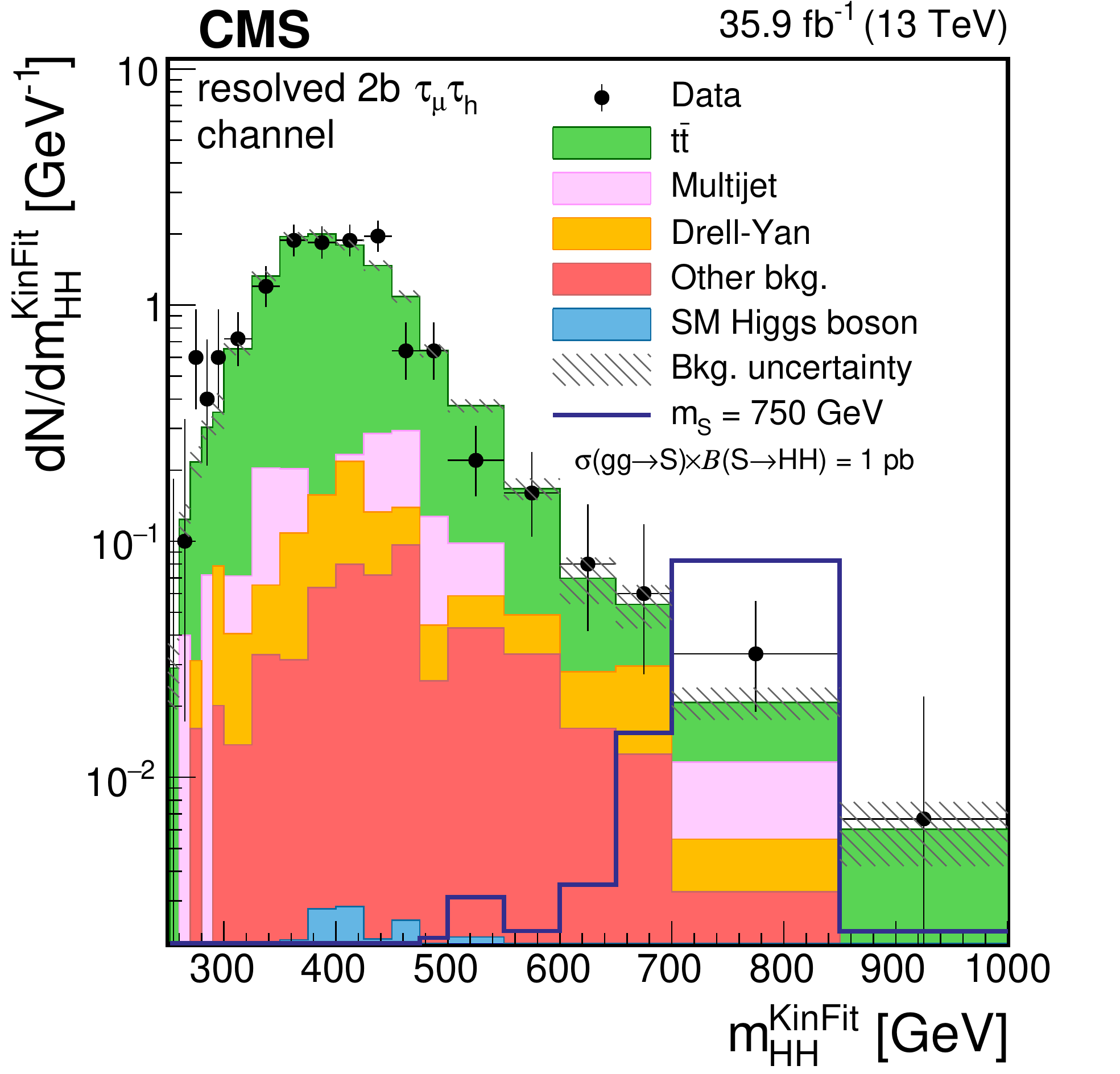}}\hspace{\fill}
{\includegraphics[width=\cmsThreeWide] {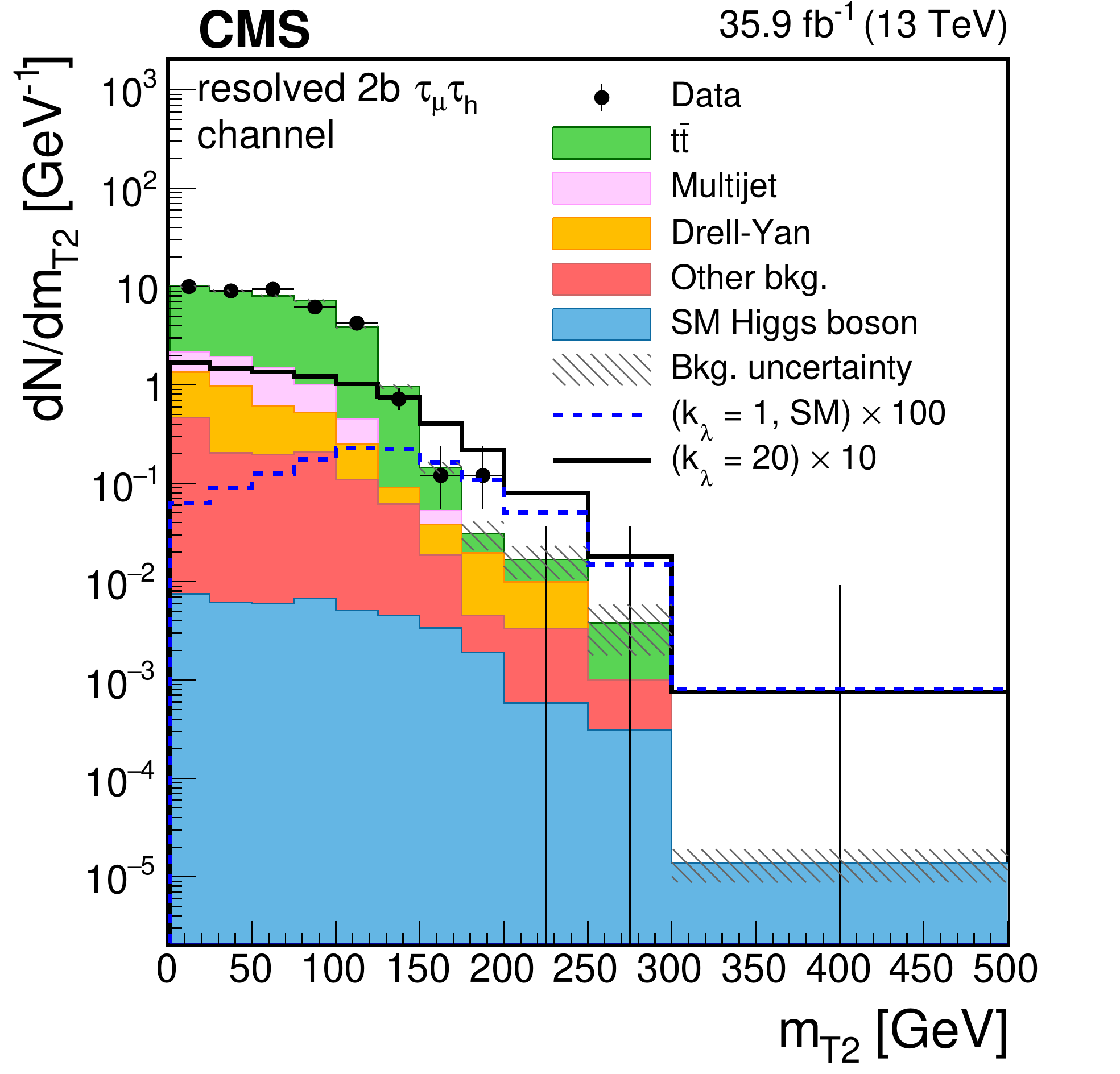}}\\
\hspace{\fill}{\includegraphics[width=\cmsThreeWide] {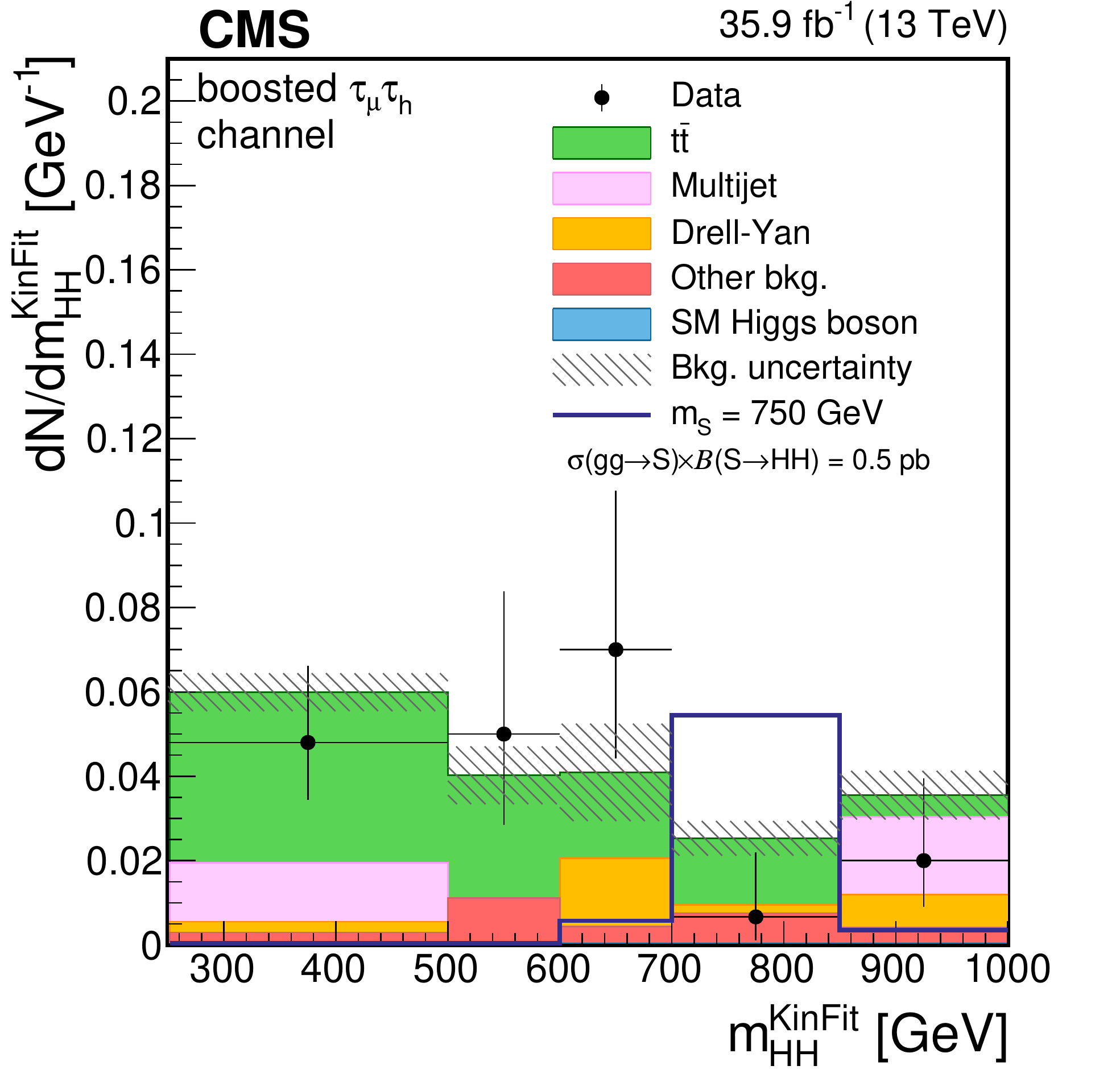}}\hspace{\fill}
{\includegraphics[width=\cmsThreeWide] {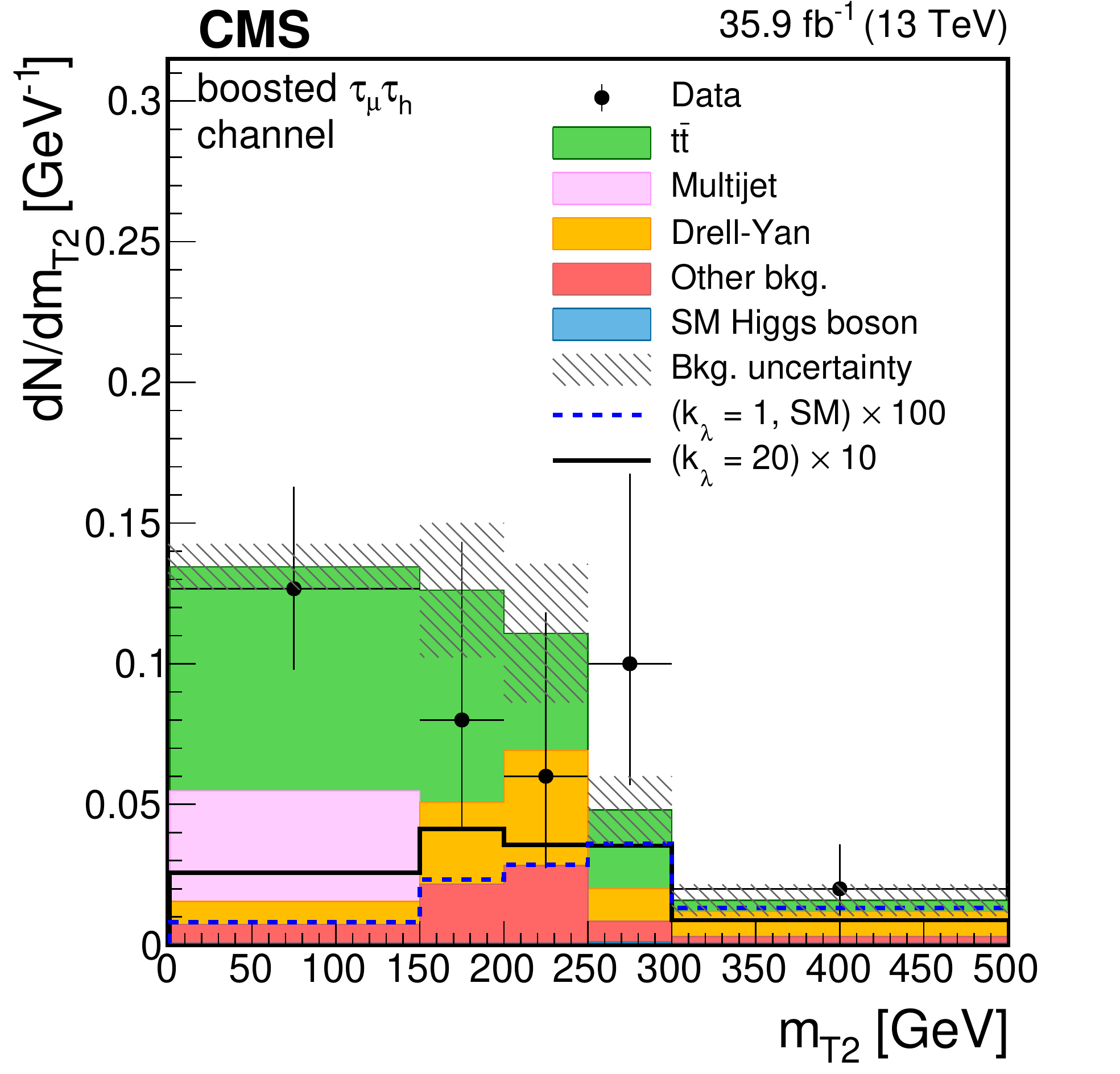}}
\caption{\label{fig:finalMuTauPlots}
Distributions of the events observed in the signal regions of the $\muth$ final state.
The first, second, and third rows show the resolved 1b1j, 2b, and boosted regions, respectively.
Panels in the
right column show the distribution of the \MTTwo variable, while the other panels show the distribution of the \MHHKinFit variable, separated in the low-mass (LM, left panels) and high-mass (HM, central panels) regions for the resolved event categories. Data are represented by points with error bars and expected signal contributions are represented by the solid (BSM \HH signals) and dashed (SM nonresonant \HH signal) lines.
Expected background contributions (shaded histograms) and associated systematic uncertainties (dashed areas) are shown as obtained after the maximum likelihood fit to the data under the background-only hypothesis.
The background histograms are stacked while the signal histograms are not stacked.
}
\end{figure*}

\begin{figure*}[p]
\centering
{\includegraphics[width=\cmsThreeWide] {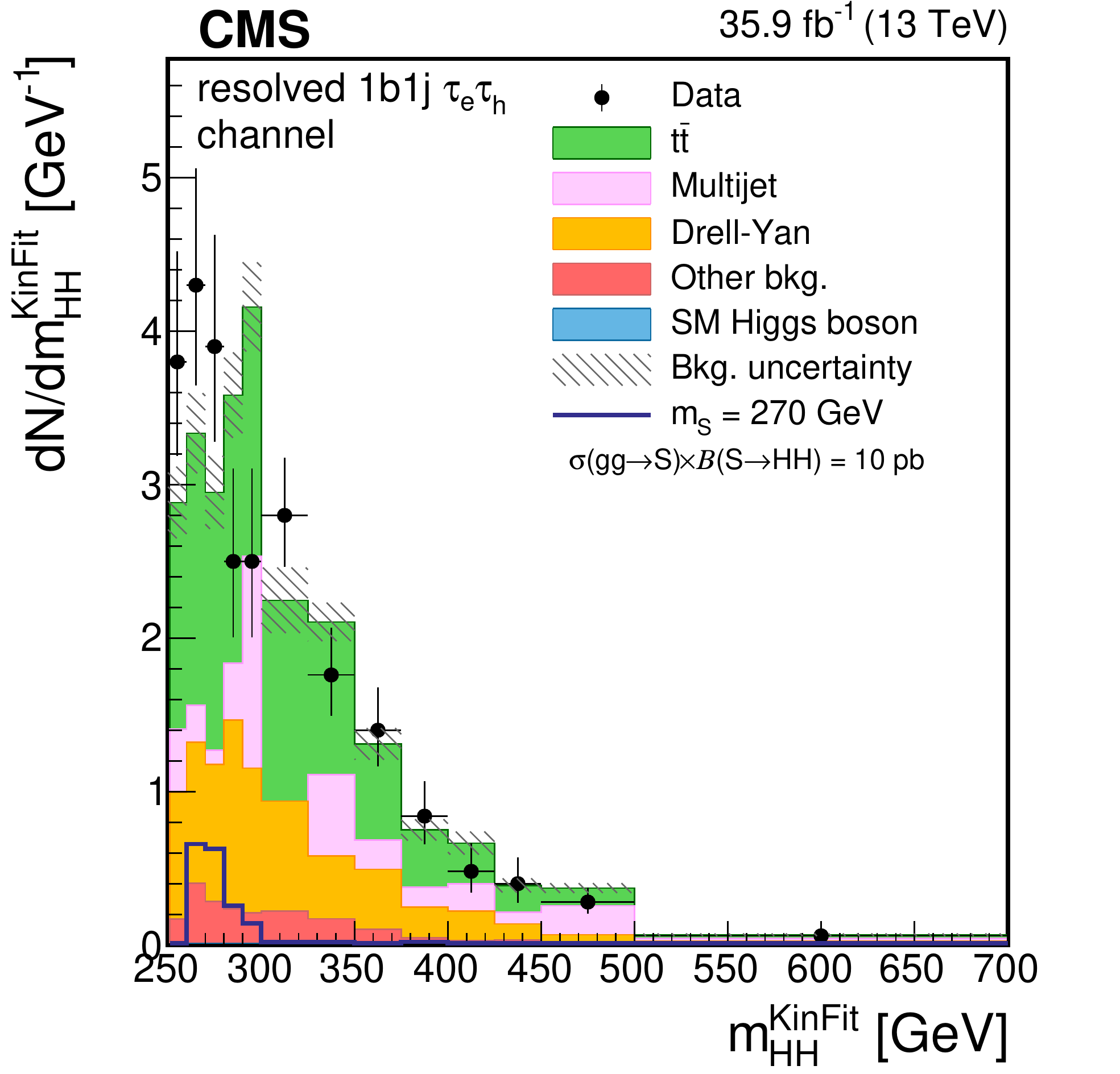}}\hspace{\fill}
{\includegraphics[width=\cmsThreeWide] {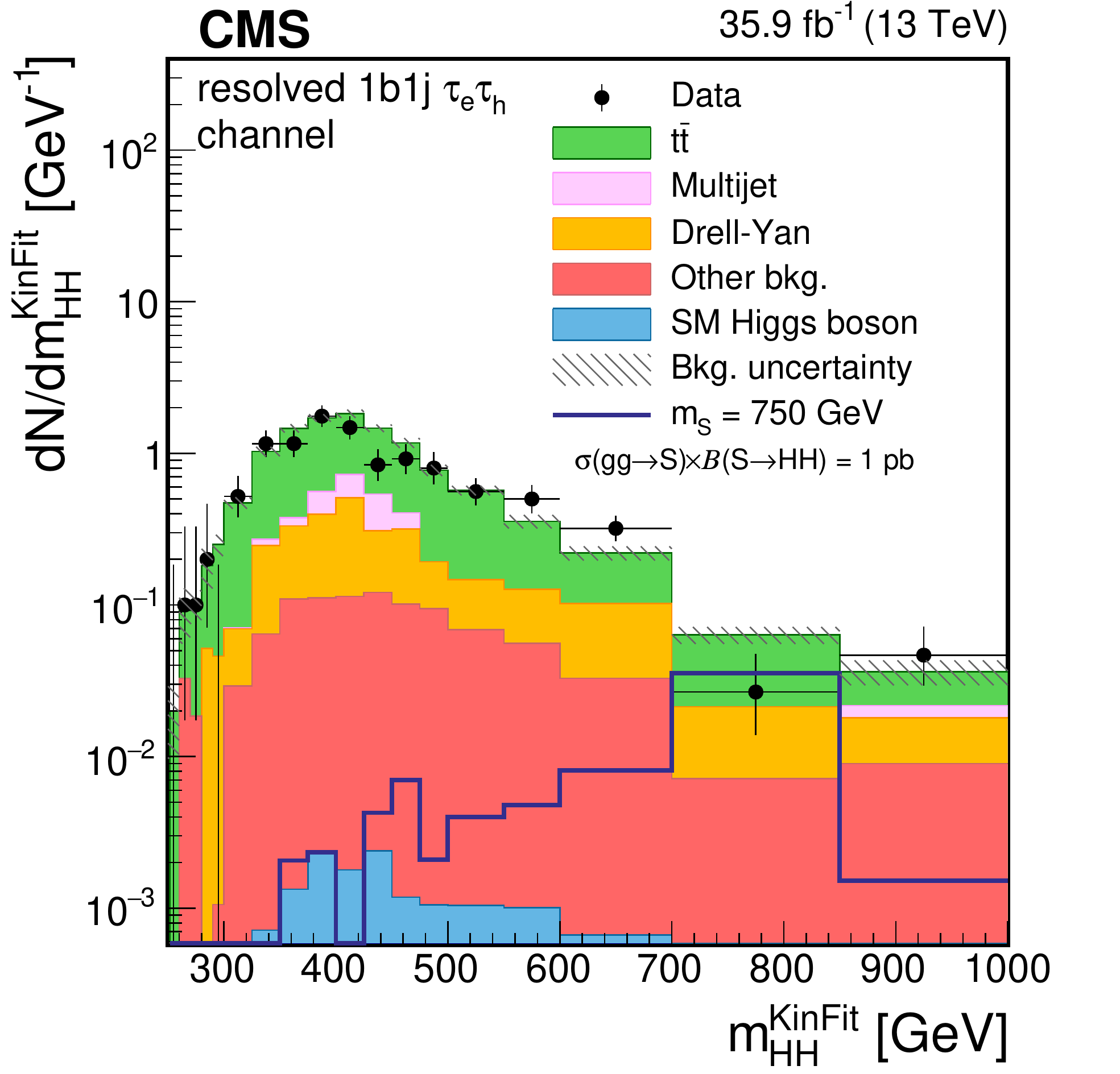}}\hspace{\fill}
{\includegraphics[width=\cmsThreeWide] {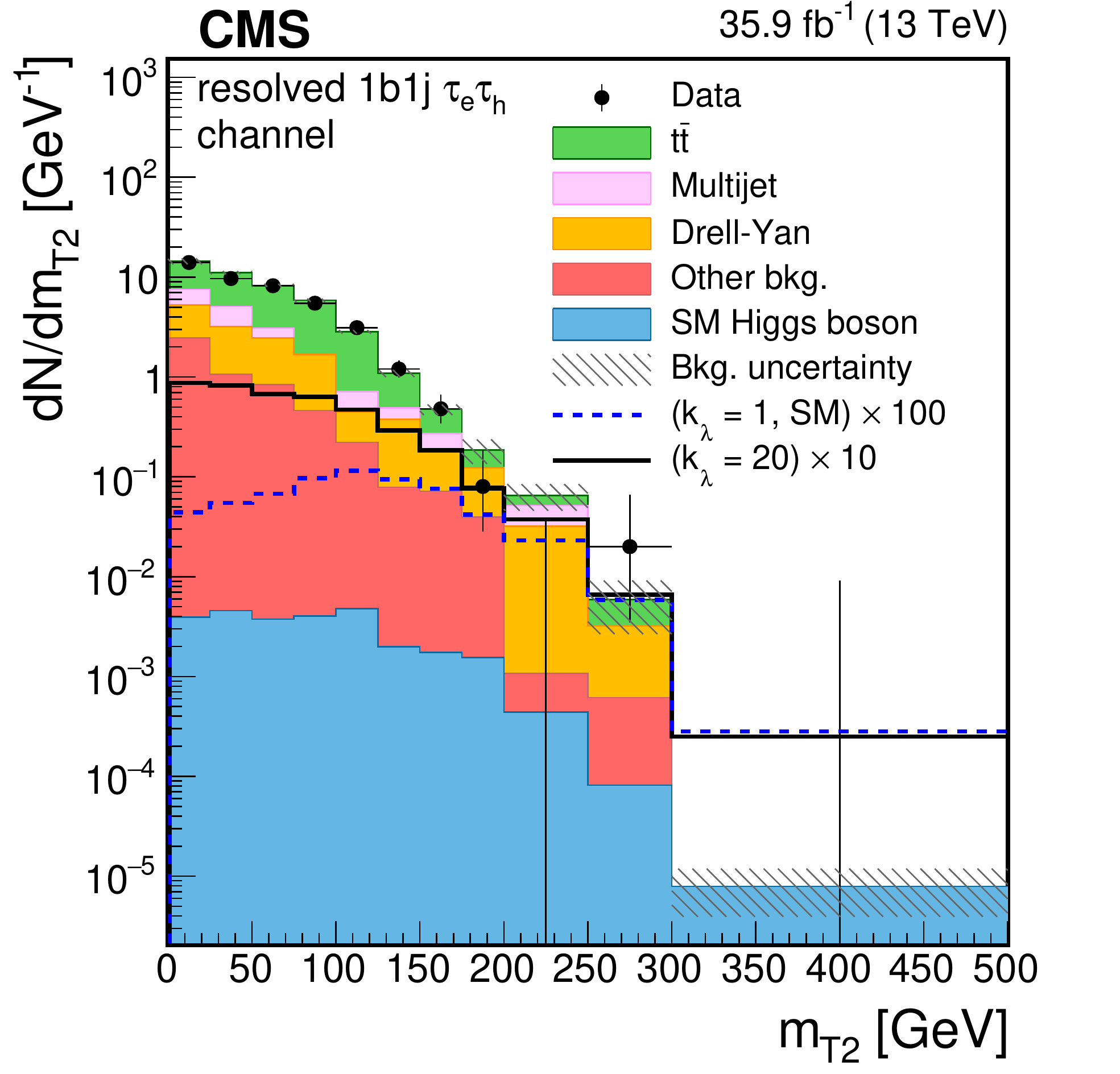}}\\
 {\includegraphics[width=\cmsThreeWide] {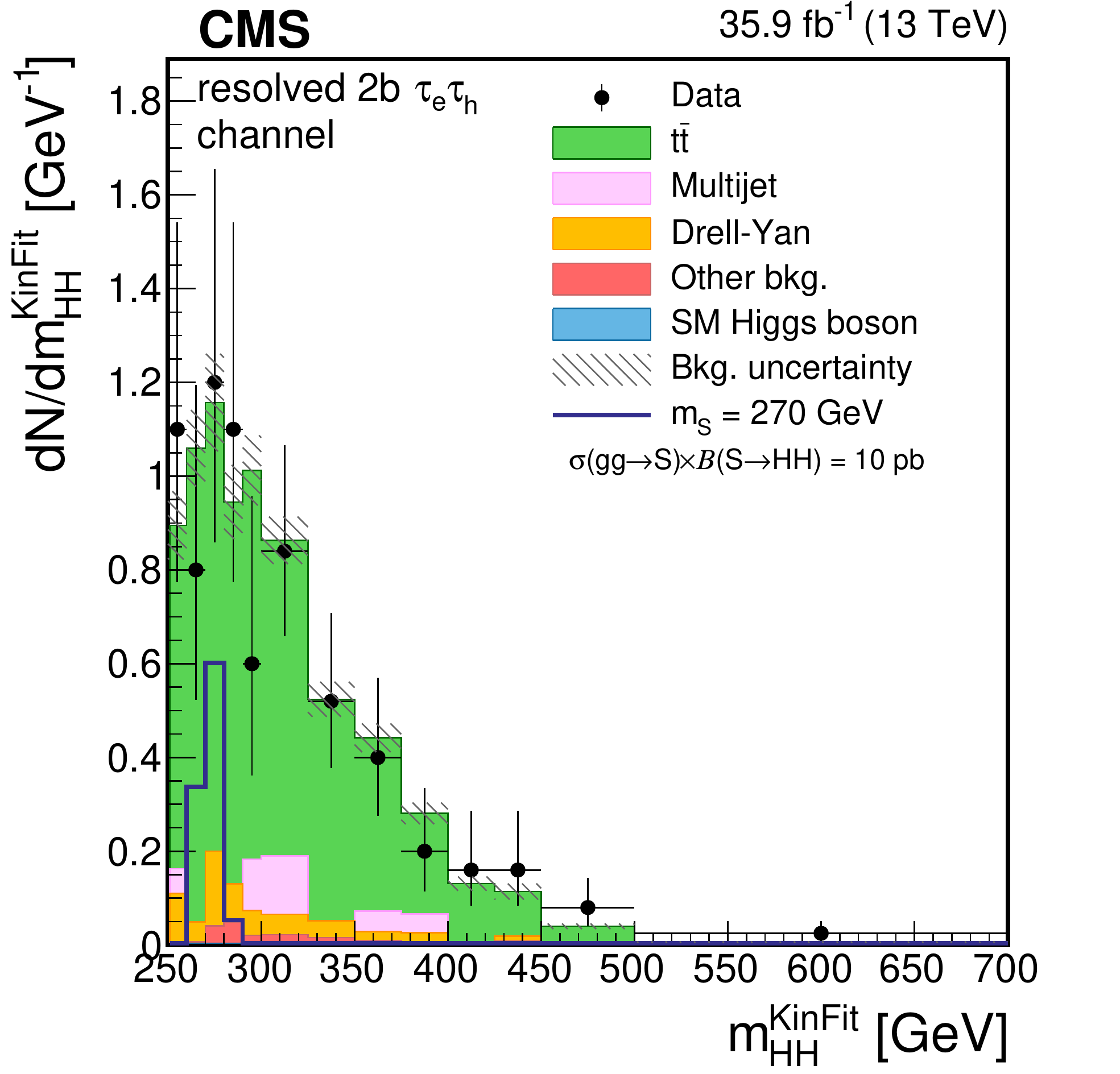}}\hspace{\fill}
 {\includegraphics[width=\cmsThreeWide] {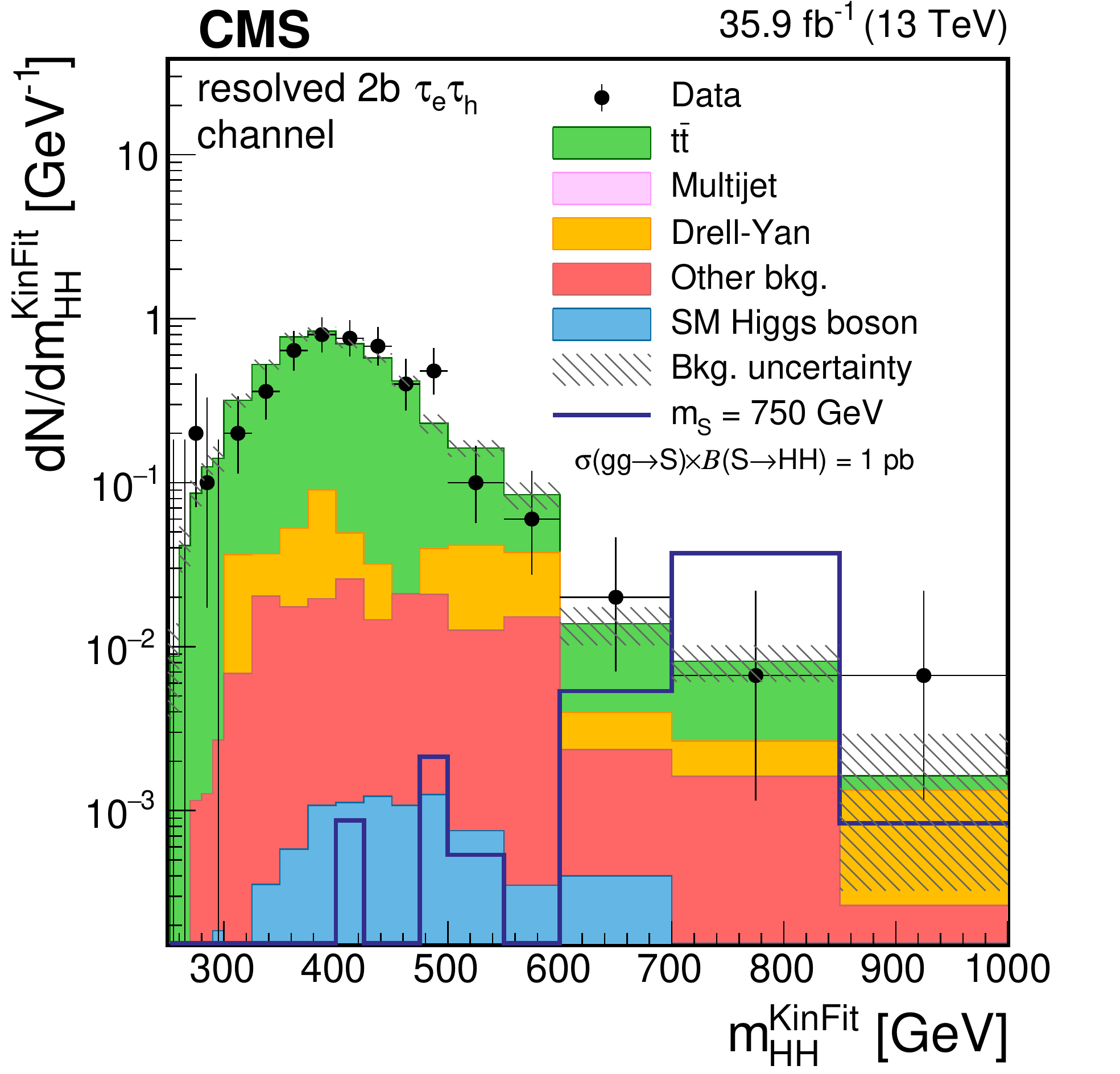}}\hspace{\fill}
{\includegraphics[width=\cmsThreeWide] {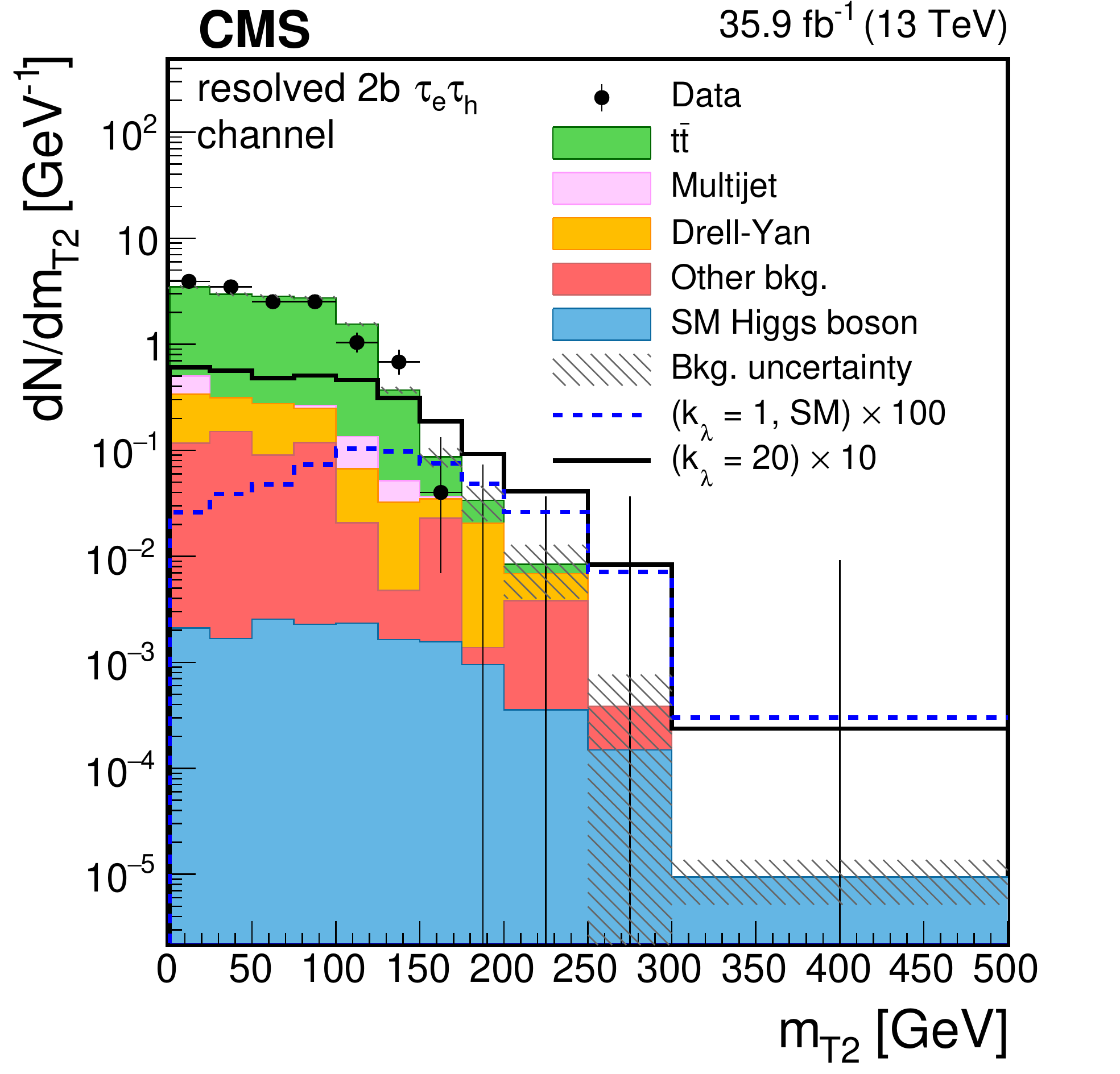}}\\
\hspace{\fill} {\includegraphics[width=\cmsThreeWide] {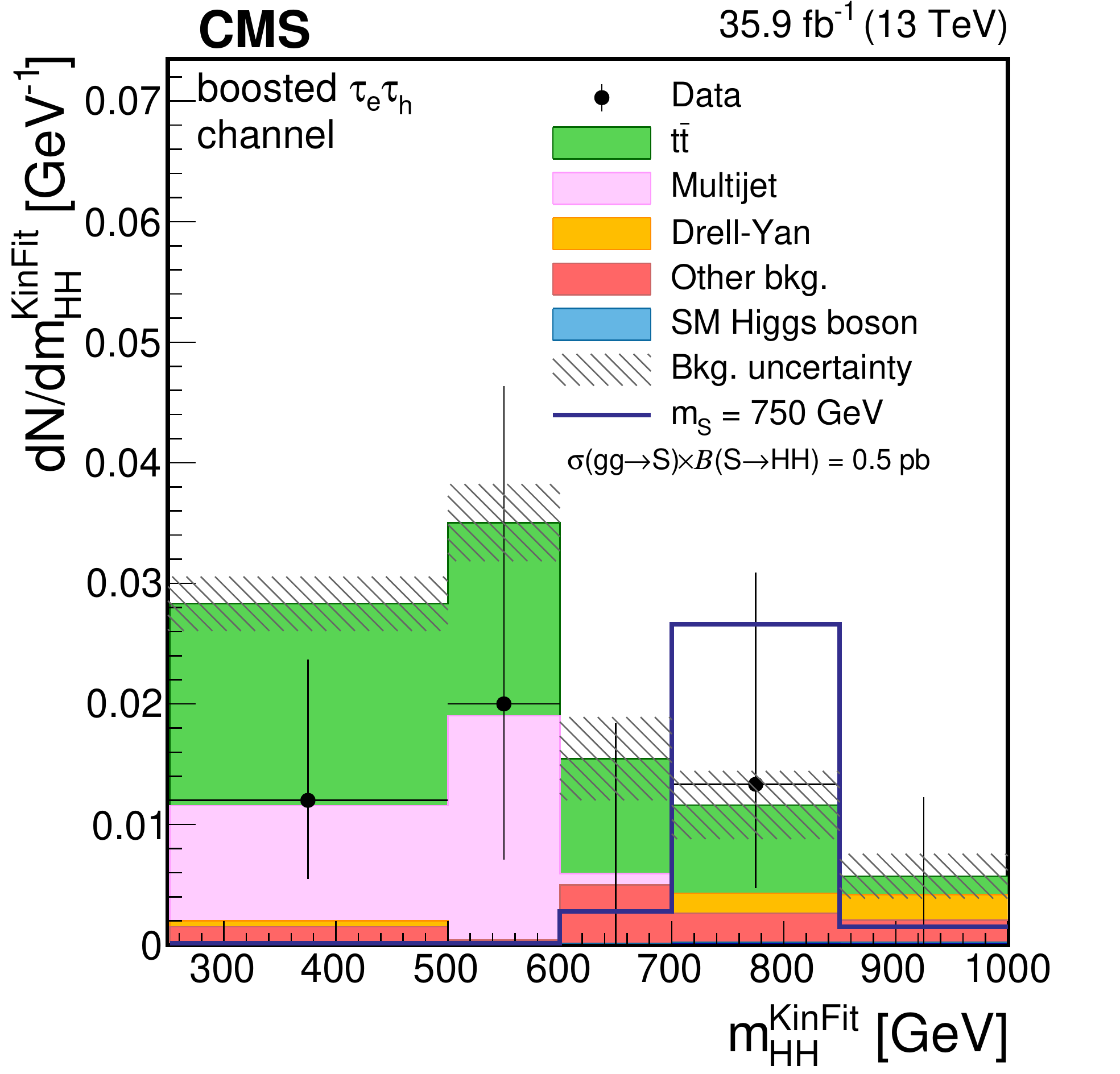}}\hspace{\fill}
{\includegraphics[width=\cmsThreeWide] {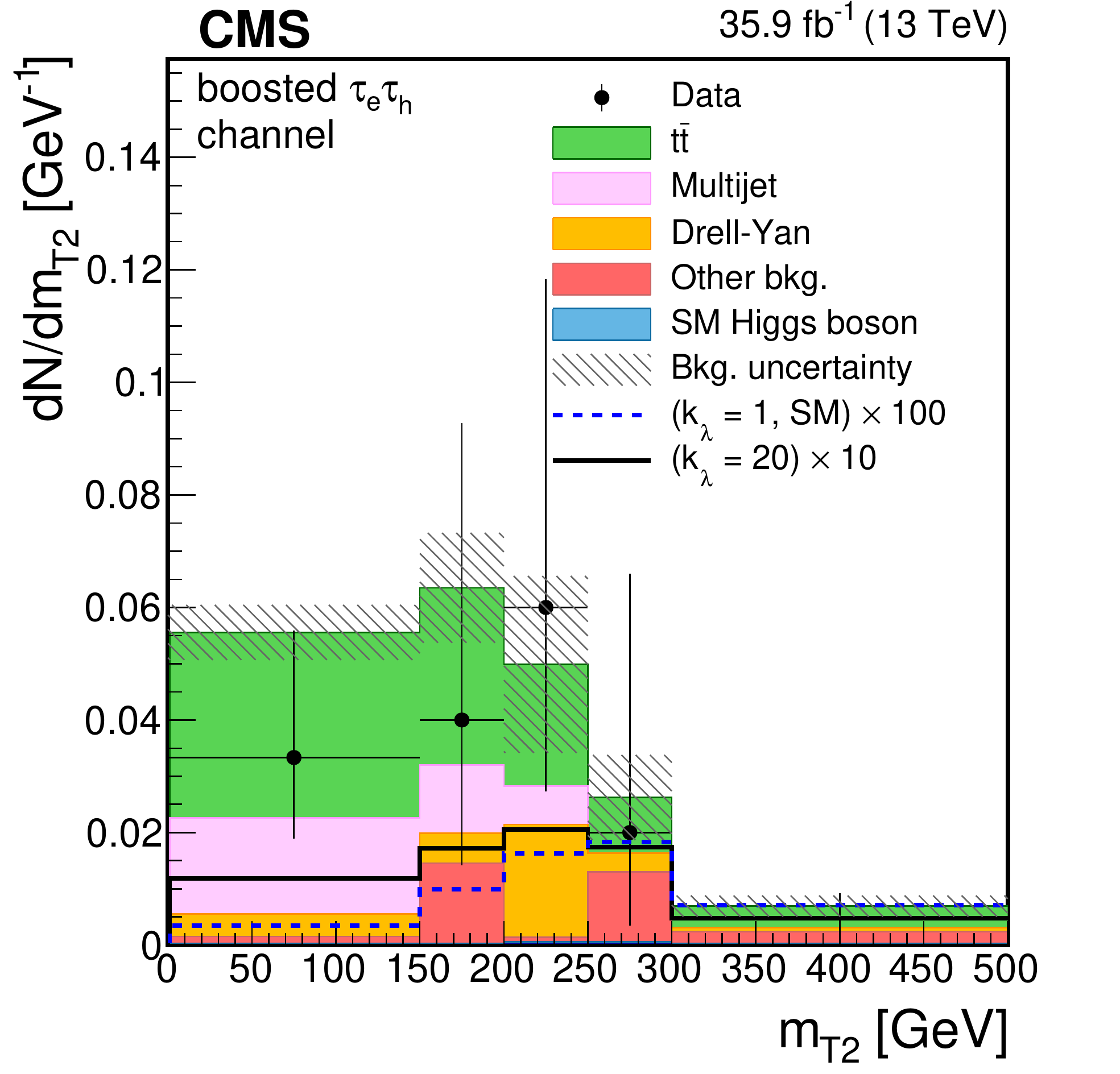}}
\caption{\label{fig:finalETauPlots}
Distributions of the events observed in the signal regions of the $\eleth$ final state.
The first, second, and third rows show the resolved 1b1j, 2b, and boosted regions, respectively.
Panels in the right column show the distribution of the \MTTwo variable, while the other panels show the distribution of the \MHHKinFit variable, separated in the low-mass (LM, left panels) and high-mass (HM, central panels) regions for the resolved event categories. Data are represented by points with error bars and expected signal contributions are represented by the solid (BSM \HH signals) and dashed (SM nonresonant \HH signal) lines.
Expected background contributions (shaded histograms) and associated systematic uncertainties (dashed areas) are shown as obtained after the maximum likelihood fit to the data under the background-only hypothesis.
The background histograms are stacked while the signal histograms are not stacked.
}
\end{figure*}

\begin{figure*}[p]
\centering
{\includegraphics[width=\cmsThreeWide] {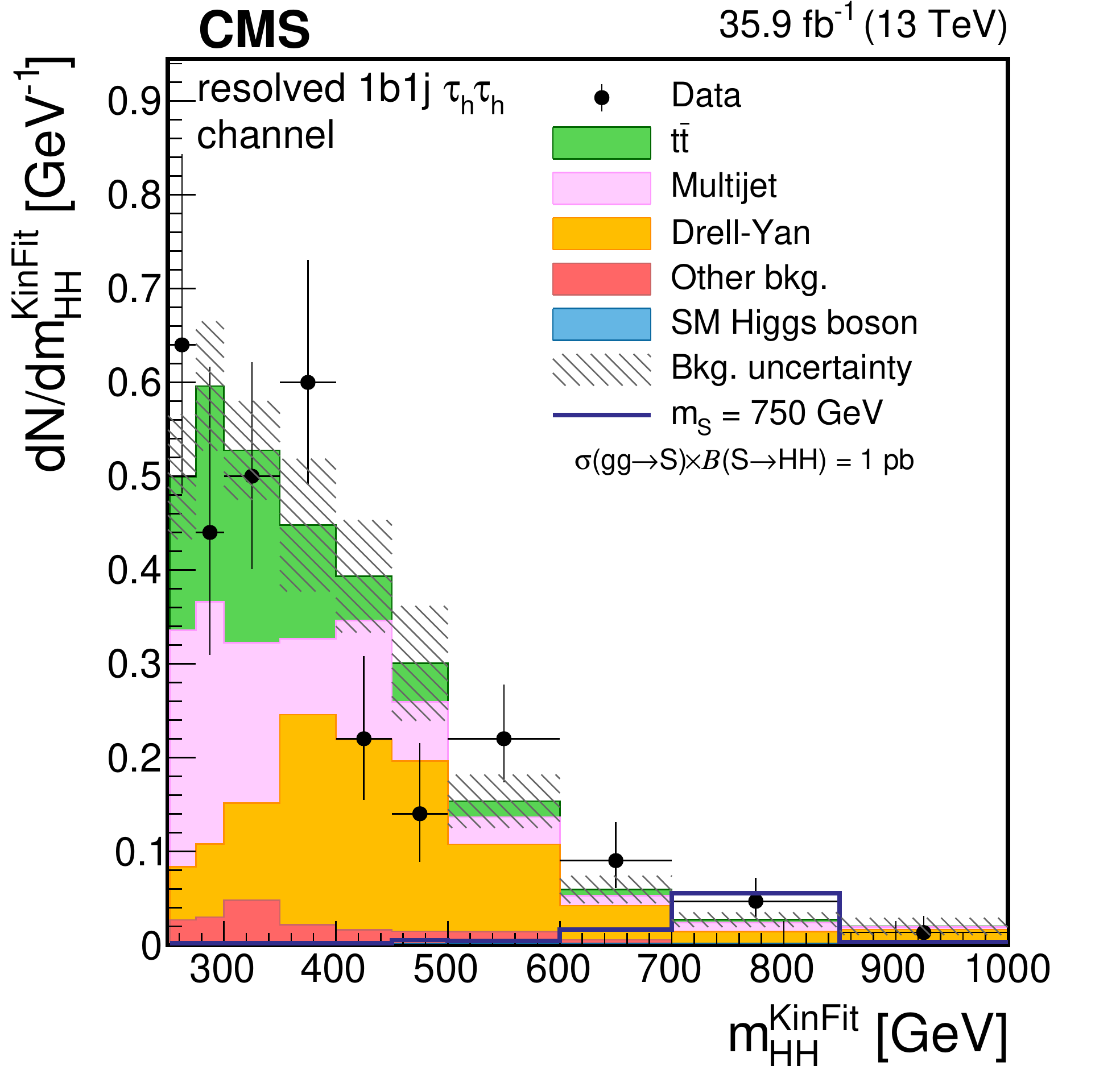}}
{\includegraphics[width=\cmsThreeWide] {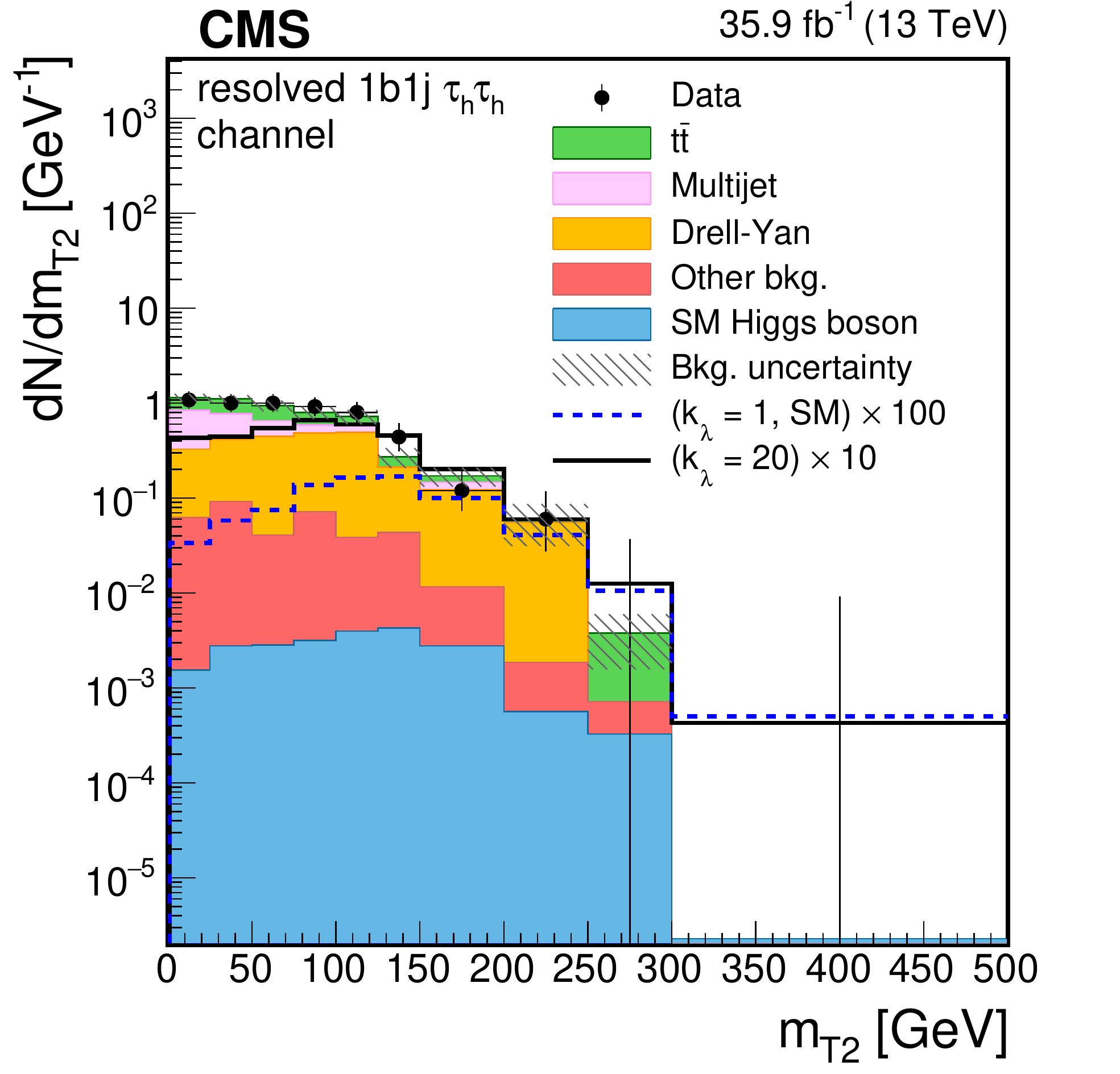}}\\
{\includegraphics[width=\cmsThreeWide] {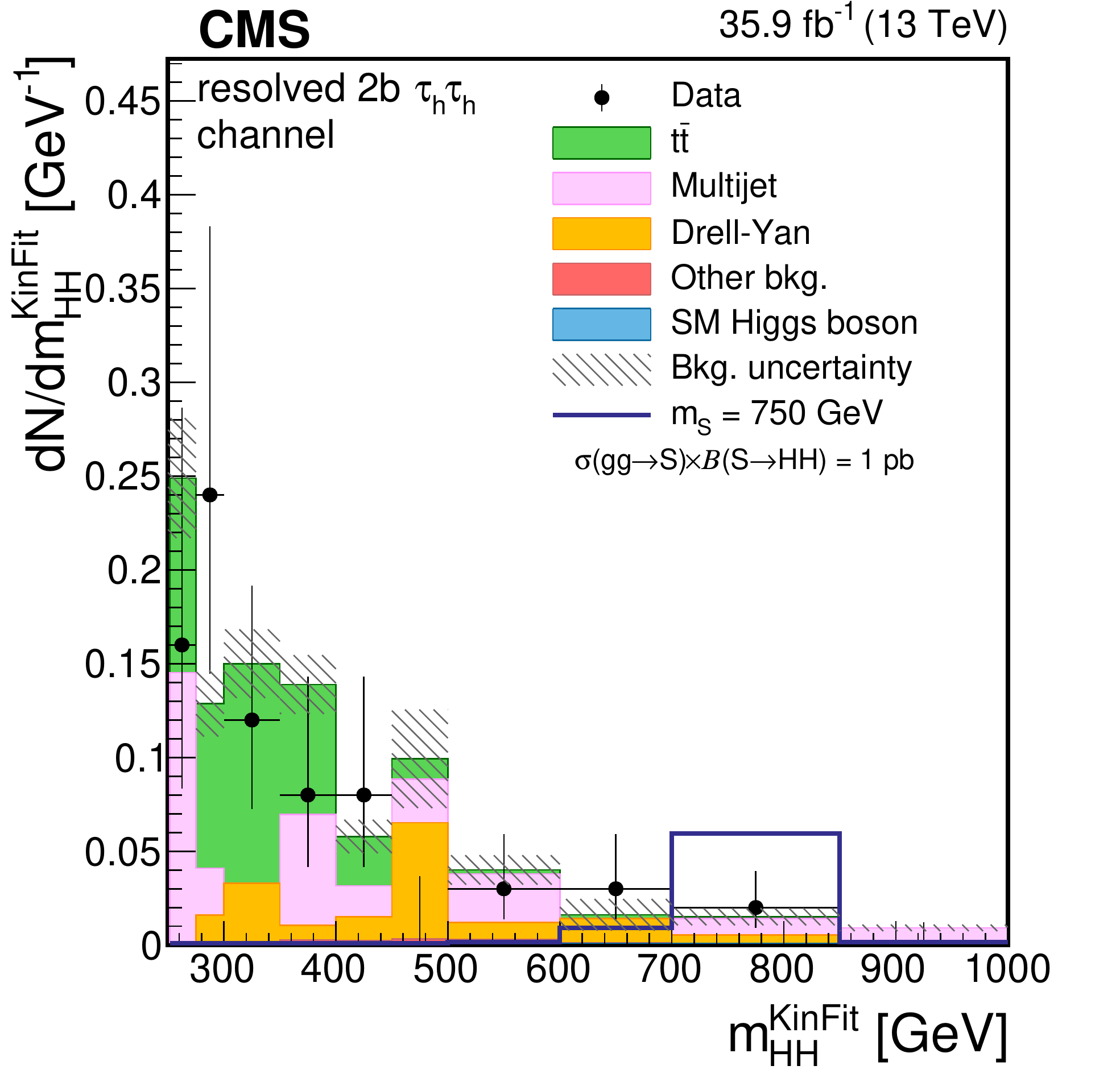}}
{\includegraphics[width=\cmsThreeWide] {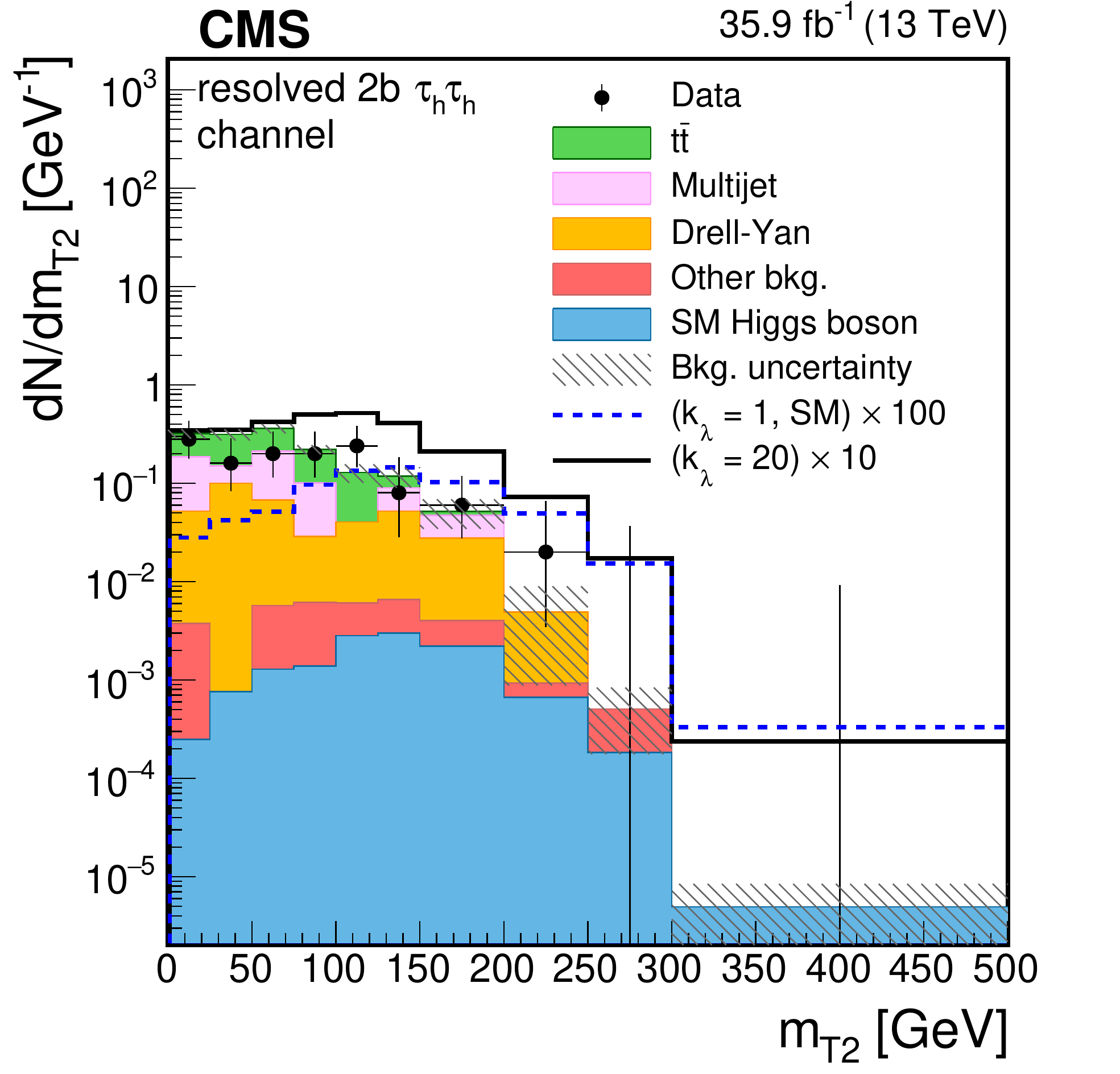}}\\
{\includegraphics[width=\cmsThreeWide] {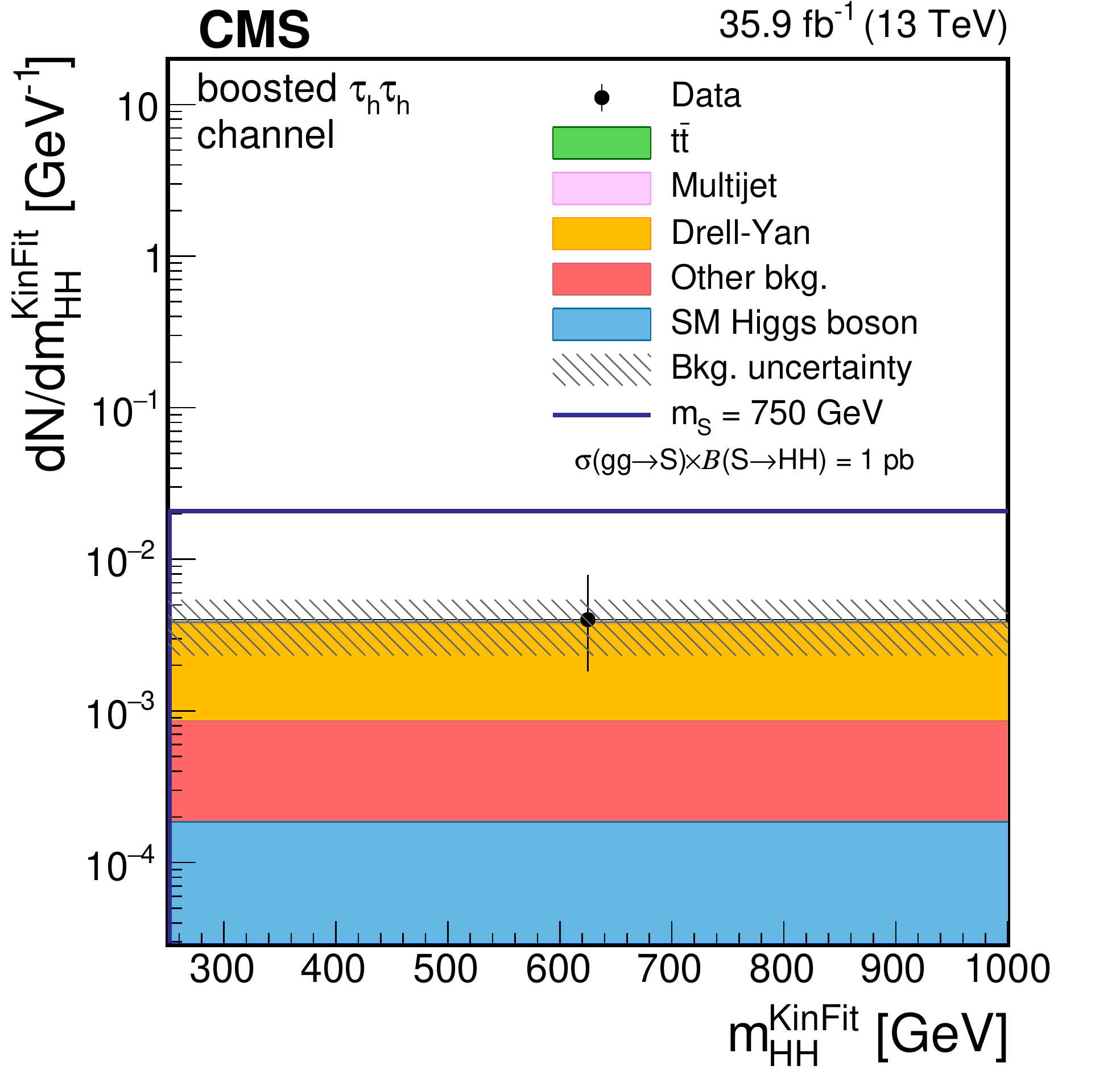}}
{\includegraphics[width=\cmsThreeWide] {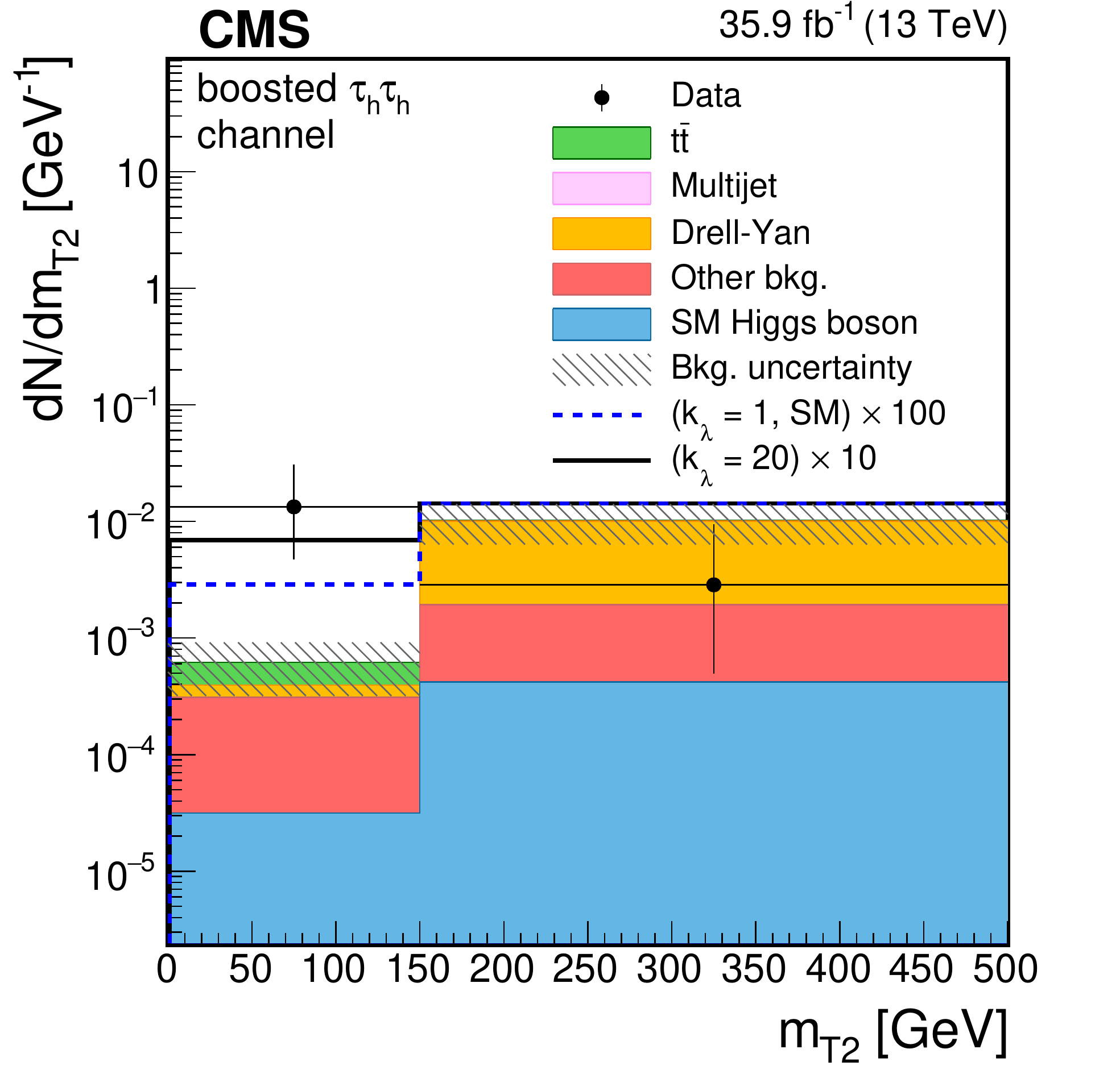}}\\
\caption{\label{fig:finalTauTauPlots}
Distributions of the events observed in the signal regions of the $\thth$ final state.
The first, second, and third rows show the resolved 1b1j, 2b, and boosted regions, respectively.
Panels in the left column
show the distribution of the \MHHKinFit variable and panels
in the right column
show the distribution of the \MTTwo variable.
Data are represented by points with error bars and expected signal contributions are represented by the solid (BSM \HH signals) and dashed (SM nonresonant \HH signal) lines.
Expected background contributions (shaded histograms) and associated systematic uncertainties (dashed areas) are shown as obtained after the maximum likelihood fit to the data under the background-only hypothesis.
The background histograms are stacked while the signal histograms are not stacked.
}
\end{figure*}

Figures \ref{fig:finalMuTauPlots}, \ref{fig:finalETauPlots}, and \ref{fig:finalTauTauPlots} show the distributions
of the \MHHKinFit and \MTTwo variables in the \muth, \eleth, and \thth final states, respectively.
The expected signature of resonant \HH production is a localized excess in the \MHHKinFit distribution, while an enhancement in the tails of the \MTTwo distribution would reveal the presence of nonresonant \HH production.
A binned maximum likelihood fit is performed simultaneously in the signal regions defined in this search for the three final states considered. The systematic uncertainties  discussed previously in Section~\ref{sec:syst} are introduced as nuisance parameters in the maximum likelihood fit.
In the absence of evidence for a signal, we set 95\% CL upper limits on the cross section for Higgs boson pair production using the asymptotic modified frequentist method (asymptotic $\text{CL}_\text{s}$)~\cite{Junk:1999kv,Read:2002hq}.

\begin{figure}[!bt]
\centering
{\includegraphics[width=\cmsFigWidth] {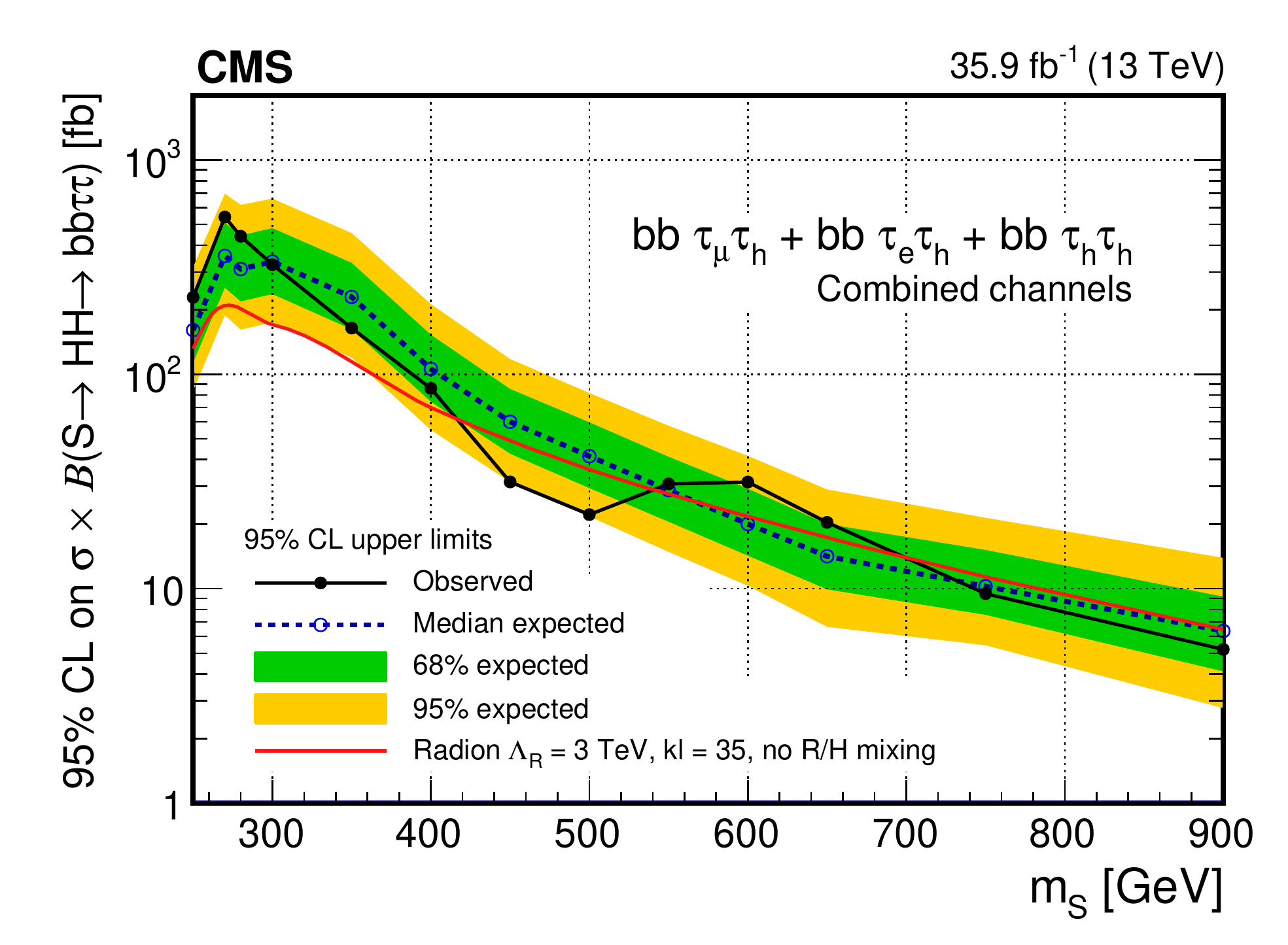}}\\
{\includegraphics[width=\cmsFigWidth] {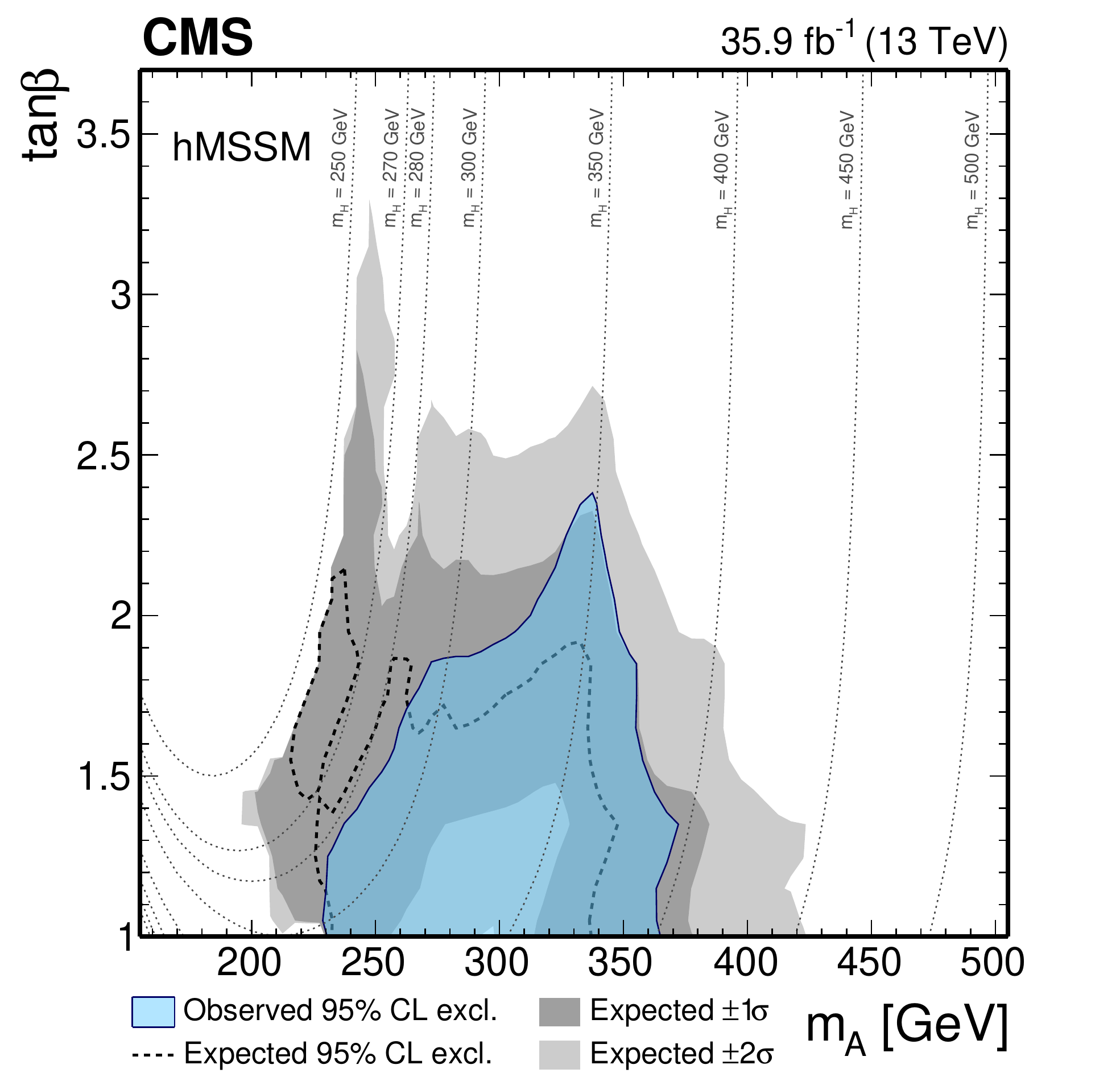}}
\caption{
\label{fig:UL_res}
(upper) Observed and expected 95\% CL upper limits on cross section times branching fraction as a function of
the mass of the resonance $\mX$ under the hypothesis that its intrinsic width is negligible with respect to the
experimental resolution. The inner (green) band and the outer (yellow) band indicate the regions containing 68 and 95\%, respectively, of the distribution of limits expected under the background-only hypothesis.
The red line denotes the expectation for the production of a radion, a spin-0 state predicted in WED models, for the parameters $\Lambda_\text{R} = 3\TeV$ (mass scale) and $\text{kl} = 35$ (size of the extra dimension), assuming the absence of mixing with the Higgs boson.
(lower) Interpretation of the exclusion limit in the context of the hMSSM model, parametrized
as a function of the $\tan \beta$ and $m_\mathrm{A}$ parameters. In this model, the CP-even lighter scalar is assumed
to be the observed 125\GeV Higgs boson and is denoted as h, while the CP-even heavier scalar is denoted as H and the CP-odd scalar is denoted as A.
The dotted lines indicate trajectories in the plane corresponding to equal values of the mass of the CP-even heavier scalar of the model, $m_\text{H}$.
}
\end{figure}

For the resonant production mode, limits are set as a function of the mass of the resonance $\mX$ under the hypothesis
that its intrinsic width is negligible compared to the experimental resolution.
The observed and expected 95\% CL limits are shown in Fig.~\ref{fig:UL_res}, upper panel.
The figure also shows the expectation for radion production, a spin-0 state predicted in WED models, for the parameters $\Lambda_\text{R} = 3\TeV$ (mass scale) and $\text{kl} = 35$ (size of the extra dimension), and assuming the absence of mixing with the Higgs boson. The corresponding cross section and branching fractions are taken from~\cite{Oliveira:2014kla}.
These model-independent limits are also interpreted in the hMSSM scenario~\cite{Djouadi:2013uqa, Djouadi:2015jea},
that is a parametrization of the MSSM that considers the observed 125\GeV Higgs boson as the lighter scalar predicted from the model (usually denoted as h in the context of the model),
while the resonance of mass $\mX$ represents the heavier CP-even scalar
(usually denoted as H in the context of the model). Excluded regions as a function of the $m_\mathrm{A}$ and $\tan\beta$ parameters, representing respectively the mass of the CP-odd scalar and the ratio of the vacuum expectation values of the two Higgs doublets of the model, are shown in Fig.~\ref{fig:UL_res}, lower panel. The minimum of the sensitivity around $\mX = 270$\GeV results in the presence of two separate expected excluded regions in this interpretation.

\begin{figure}[!bt]
\centering
{\includegraphics[width=\cmsFigWidth] {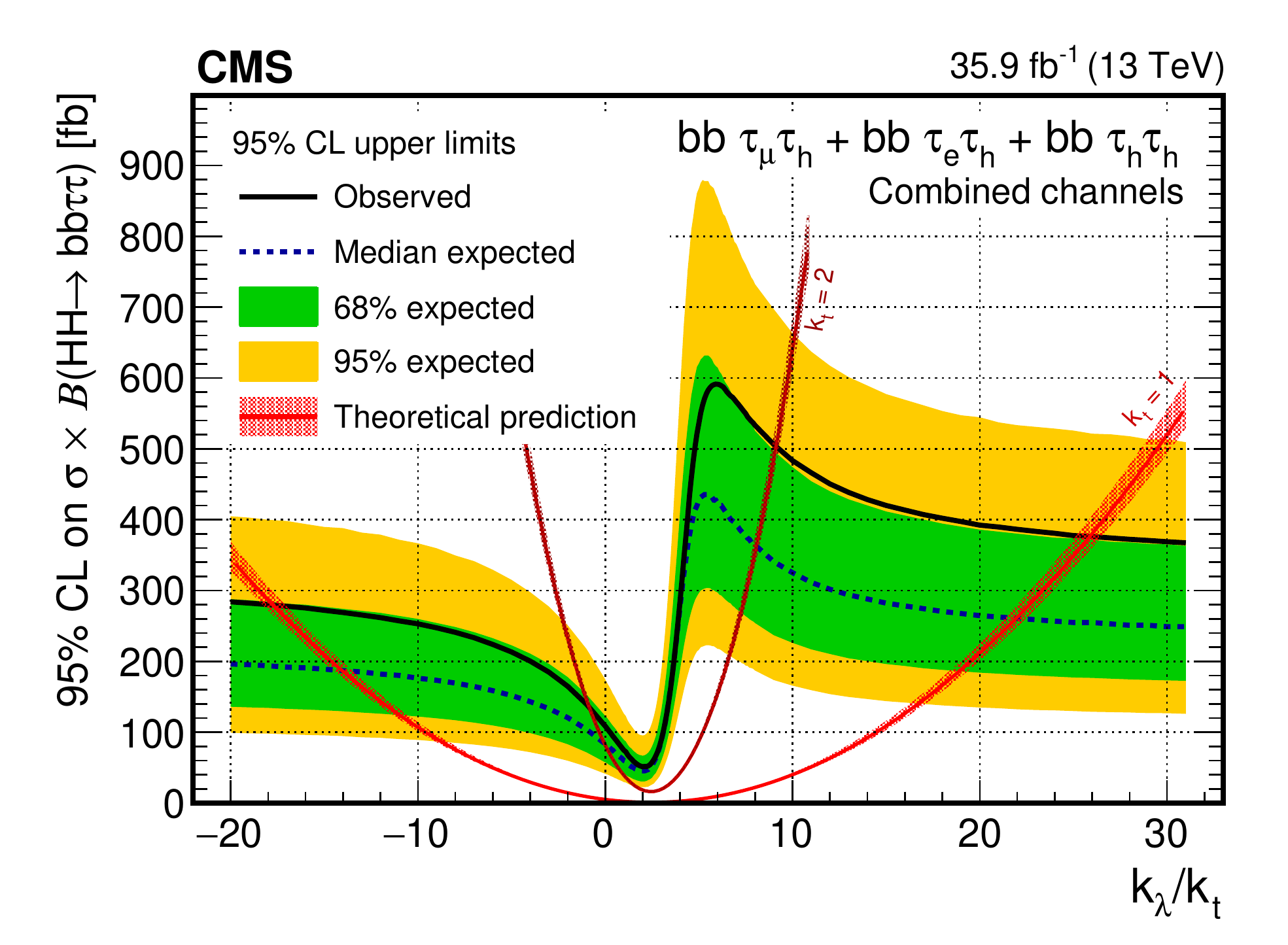}}\\
{\includegraphics[width=\cmsFigWidth] {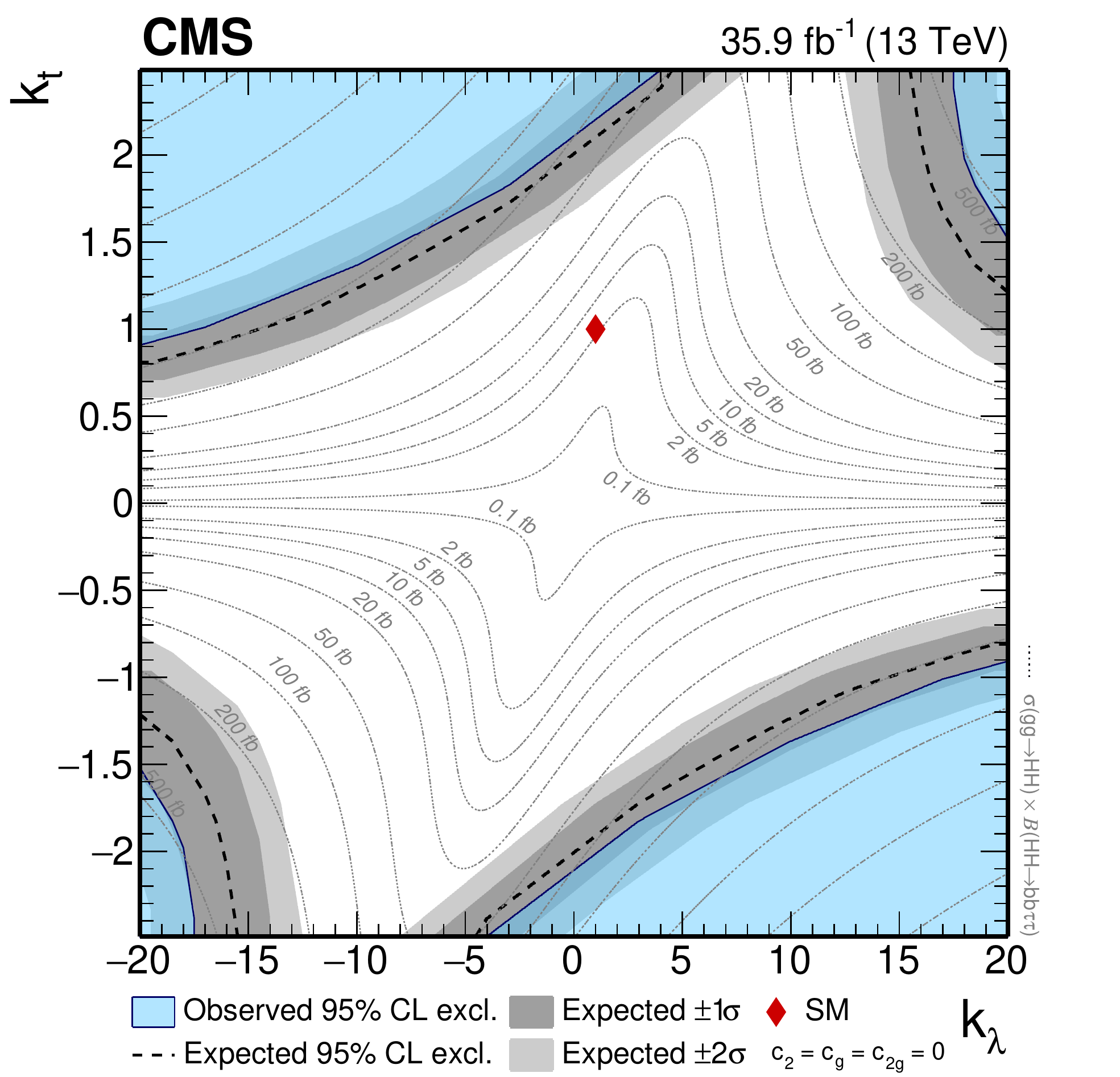}}
\caption{
\label{fig:UL_nonres}
(upper) Observed and expected 95\% CL upper limits on cross section times branching fraction as a function of
$k_\lambda/k_\PQt$. The inner (green) band and the outer (yellow) band indicate the regions containing 68 and 95\%, respectively, of the distribution of limits expected under the background-only hypothesis. The two red bands show the theoretical cross section expectations and the corresponding uncertainties for $k_\PQt = 1$ and $k_\PQt = 2$.
(lower) Test of $k_\lambda$ and $k_\PQt$ anomalous couplings. The blue region denotes the parameters excluded
by the data at 95\% CL, while the dashed black line and the grey regions denote the expected exclusions and the $1\sigma$
and $2\sigma$ bands.
The dotted lines indicate trajectories in the plane with equal values of cross section times branching fraction that are displayed in the associated labels. The diamond-shaped symbol denotes the couplings predicted by the SM.
The theory predictions and the expected and observed limits are symmetric through a $(k_\lambda$, $k_\PQt) \leftrightarrow (-k_\lambda, -k_\PQt)$ transformation.
In both figures, the couplings that are not explicitly tested are assumed to correspond to the SM prediction.
}
\end{figure}

For the nonresonant production mode, including the theoretical uncertainties, the observed 95\% CL upper limit on the \HH production cross section times branching fraction amounts to 75.4\fb while the expected 95\% CL upper limit amounts to 61.0\unit{fb}. These values correspond to about 30 and 25 times the SM prediction, respectively.
Limits are set for different hypotheses of anomalous self-coupling and top quark coupling of the Higgs boson.
The signal kinematics depend on the ratio of the two couplings and 95\% CL upper limits are set as a function
of $k_\lambda/k_\PQt$, assuming the other BSM couplings to be zero. The result is shown in Fig.~\ref{fig:UL_nonres}, upper panel,
and the exclusion is compared with the theoretical prediction for the cross section for $k_\PQt = 1$ and $k_\PQt = 2$.
The sensitivity varies as a function of $k_\lambda$ and $k_\PQt$ because of the corresponding changes in the signal \MTTwo distribution.
These upper limits are used to set constraints on anomalous $k_\lambda$ and $k_\PQt$ couplings as shown in Fig.~\ref{fig:UL_nonres}, lower panel, where the $\text{c}_2$, $\text{c}_{2\text{g}}$, and $\text{c}_\text{g}$ couplings are assumed to be equal to zero. The branching fractions for the decays of the Higgs boson into a $\PQb\PQb$ and
$\PGt\PGt$ pair are assumed to be those predicted by the SM for all the values of $k_\lambda$ and $k_\PQt$ tested.

\section{Summary}

A search for resonant and nonresonant Higgs boson pair (\HH) production in the \bbtt final state is presented.
This search uses a data sample collected in proton-proton collisions at $\sqrt{s} = 13 \TeV$ that corresponds to an integrated luminosity of $35.9\fbinv$. The three most sensitive decay channels of the \PGt lepton pair, requiring the decay of one or both
$\PGt$ leptons into final-state hadrons and a neutrino, are used.
The results are found to be statistically compatible with the expected standard model (SM) background contribution,
and upper limits at the 95\% confidence level are set on the \HH production cross sections.

For the resonant production mechanism, upper exclusion limits at 95\% confidence level (CL) are obtained for the production of a narrow resonance
of mass $\mX$ ranging from 250 to 900\GeV. These model-independent results are interpreted in the context of the hMSSM scenario,
where a region in the parameter space corresponding to values of $m_\mathrm{A}$ between 230 and 360\GeV and $\tan\beta\lesssim2$ is
excluded at 95\% CL.

For the nonresonant production mechanism, the theoretical framework of an
effective Lagrangian is used to parametrize the cross section as a function of anomalous couplings of the Higgs boson.
Upper limits at 95\% CL on the \HH cross section are obtained as a function of
$k_\lambda = \lambdahhh/\lambdahhh^\text{SM}$ and $k_\PQt = y_\PQt/y^\text{SM}_\PQt$.
The observed 95\% CL upper limit corresponds to approximately 30 times the theoretical prediction for the SM cross section, and the expected limit is about 25 times the SM prediction.
This is the highest sensitivity achieved so far for SM \HH production at the LHC.

\begin{acknowledgments}
We congratulate our colleagues in the CERN accelerator departments for the excellent performance of the LHC and thank the technical and administrative staffs at CERN and at other CMS institutes for their contributions to the success of the CMS effort. In addition, we gratefully acknowledge the computing centres and personnel of the Worldwide LHC Computing Grid for delivering so effectively the computing infrastructure essential to our analyses. Finally, we acknowledge the enduring support for the construction and operation of the LHC and the CMS detector provided by the following funding agencies: BMWFW and FWF (Austria); FNRS and FWO (Belgium); CNPq, CAPES, FAPERJ, and FAPESP (Brazil); MES (Bulgaria); CERN; CAS, MoST, and NSFC (China); COLCIENCIAS (Colombia); MSES and CSF (Croatia); RPF (Cyprus); SENESCYT (Ecuador); MoER, ERC IUT, and ERDF (Estonia); Academy of Finland, MEC, and HIP (Finland); CEA and CNRS/IN2P3 (France); BMBF, DFG, and HGF (Germany); GSRT (Greece); OTKA and NIH (Hungary); DAE and DST (India); IPM (Iran); SFI (Ireland); INFN (Italy); MSIP and NRF (Republic of Korea); LAS (Lithuania); MOE and UM (Malaysia); BUAP, CINVESTAV, CONACYT, LNS, SEP, and UASLP-FAI (Mexico); MBIE (New Zealand); PAEC (Pakistan); MSHE and NSC (Poland); FCT (Portugal); JINR (Dubna); MON, RosAtom, RAS, RFBR and RAEP (Russia); MESTD (Serbia); SEIDI, CPAN, PCTI and FEDER (Spain); Swiss Funding Agencies (Switzerland); MST (Taipei); ThEPCenter, IPST, STAR, and NSTDA (Thailand); TUBITAK and TAEK (Turkey); NASU and SFFR (Ukraine); STFC (United Kingdom); DOE and NSF (USA).

\hyphenation{Rachada-pisek} Individuals have received support from the Marie-Curie programme and the European Research Council and Horizon 2020 Grant, contract No. 675440 (European Union); the Leventis Foundation; the A. P. Sloan Foundation; the Alexander von Humboldt Foundation; the Belgian Federal Science Policy Office; the Fonds pour la Formation \`a la Recherche dans l'Industrie et dans l'Agriculture (FRIA-Belgium); the Agentschap voor Innovatie door Wetenschap en Technologie (IWT-Belgium); the Ministry of Education, Youth and Sports (MEYS) of the Czech Republic; the Council of Science and Industrial Research, India; the HOMING PLUS programme of the Foundation for Polish Science, cofinanced from European Union, Regional Development Fund, the Mobility Plus programme of the Ministry of Science and Higher Education, the National Science Center (Poland), contracts Harmonia 2014/14/M/ST2/00428, Opus 2014/13/B/ST2/02543, 2014/15/B/ST2/03998, and 2015/19/B/ST2/02861, Sonata-bis 2012/07/E/ST2/01406; the National Priorities Research Program by Qatar National Research Fund; the Programa Clar\'in-COFUND del Principado de Asturias; the Thalis and Aristeia programmes cofinanced by EU-ESF and the Greek NSRF; the Rachadapisek Sompot Fund for Postdoctoral Fellowship, Chulalongkorn University and the Chulalongkorn Academic into Its 2nd Century Project Advancement Project (Thailand); and the Welch Foundation, contract C-1845.
\end{acknowledgments}

\bibliography{auto_generated}

\cleardoublepage \appendix\section{The CMS Collaboration \label{app:collab}}\begin{sloppypar}\hyphenpenalty=5000\widowpenalty=500\clubpenalty=5000\input{HIG-17-002-authorlist.tex}\end{sloppypar}
\end{document}

%% file: HIG-17-002-authorlist.tex
\textbf{Yerevan Physics Institute,  Yerevan,  Armenia}\\*[0pt]
A.M.~Sirunyan, A.~Tumasyan
\vskip\cmsinstskip
\textbf{Institut f\"{u}r Hochenergiephysik,  Wien,  Austria}\\*[0pt]
W.~Adam, F.~Ambrogi, E.~Asilar, T.~Bergauer, J.~Brandstetter, E.~Brondolin, M.~Dragicevic, J.~Er\"{o}, M.~Flechl, M.~Friedl, R.~Fr\"{u}hwirth\cmsAuthorMark{1}, V.M.~Ghete, J.~Grossmann, J.~Hrubec, M.~Jeitler\cmsAuthorMark{1}, A.~K\"{o}nig, N.~Krammer, I.~Kr\"{a}tschmer, D.~Liko, T.~Madlener, I.~Mikulec, E.~Pree, D.~Rabady, N.~Rad, H.~Rohringer, J.~Schieck\cmsAuthorMark{1}, R.~Sch\"{o}fbeck, M.~Spanring, D.~Spitzbart, J.~Strauss, W.~Waltenberger, J.~Wittmann, C.-E.~Wulz\cmsAuthorMark{1}, M.~Zarucki
% ...
\vskip\cmsinstskip
% ...
1:~~Also at Vienna University of Technology, Vienna, Austria\\
% ...
